\documentstyle[twocolumn,aas2pp4]{article}
\def\stacksymbols #1#2#3#4{\def\theguybelow{#2}
        \def\verticalposition{\lower#3pt}
        \def\spacingwithinsymbol{\baselineskip0pt\lineskip#4pt}
        \mathrel{\mathpalette\intermediary#1}}
\def\intermediary #1#2{\verticalposition\vbox{\spacingwithinsymbol
        \everycr={}\tabskip0pt
        \halign{$\mathsurround0pt#1\hfil##\hfil$\crcr#2\crcr
                \theguybelow\crcr}}}
\def\lta{\stacksymbols{<}{\sim}{2.5}{.2}}
\def\gta{\stacksymbols{>}{\sim}{3}{.5}}

\begin{document}
%

\title{FORMATION OF LOW MASS STARS IN ELLIPTICAL GALAXY COOLING FLOWS}

\author{William G. Mathews$^2$ and Fabrizio Brighenti$^{2,3}$}

\affil{$^2$University of California Observatories/Lick Observatory,
Board of Studies in Astronomy and Astrophysics,
University of California, Santa Cruz, CA 95064\\
mathews@lick.ucsc.edu}

\affil{$^3$Dipartimento di Astronomia,
Universit\`a di Bologna,
via Zamboni 33,
Bologna 40126, Italy\\
brighenti@astbo3.bo.astro.it}






\vskip .2in

\begin{abstract}

Thermal X-ray emission from cooling flows in elliptical 
galaxies indicates that $\sim 1$ $M_{\odot}$ of hot 
($T \sim 10^7$ K) interstellar gas cools each year, accumulating
$\sim 10^{10}$ $M_{\odot}$ over a Hubble time. 
Paradoxically, optical and 
radio frequency emission from the cooled gas is lacking, 
indicating that less than 
$\sim 10^{-3}$ of the cooled gas remains. 
Many have speculated that 
the cooled gas has formed into 
relatively invisible low mass stars, 
particularly in the context of massive cooling flows in 
galaxy clusters.
We focus here on cooling flows 
in elliptical galaxies like NGC 4472 where the cooled gas is 
made visible in emission lines from 
HII regions ionized and heated ($T_{HII} \sim 10^4$ K) 
by stellar ultraviolet radiation.
The low filling factor of HII gas requires that the 
hot gas cools at $\sim 10^6$ cooling sites within 
several kpc of the galactic center.
HII mass slowly increases at each site at
$\sim 10^{-6}$ $M_{\odot}$ yr$^{-1}$
until a neutral core develops.
Neutral cores are heated ($T_{HI} \sim 15$ K) 
and ionized ($x \sim 10^{-6}$) by thermal X-rays 
from the entire interstellar cooling flow.
We show that the {\it maximum} mass of 
spherical HI cores that become 
gravitationally unstable is only $\sim 2$ $M_{\odot}$. 
No star can exceed this mass and 
fragmentation of collapsing cores produces stars 
of even lower mass.
By this means we establish with some confidence 
that the hypothesis of low mass star formation 
is indeed correct -- the IMF is bottom heavy, 
but may be optically luminous. 
Slightly more massive stars $\lta 4.5$ $M_{\odot}$ 
can form near the effective radius 
($r = 8.57$ kpc in NGC 4472) if sufficient 
masses of interstellar gas cool there, 
producing a luminous population of 
intermediate mass stars  
perhaps with radial orbits 
that may contribute to the stellar H$\beta$ index. 
The degree of ionization in 
gravitationally collapsing 
cores is sufficiently low to allow magnetic fields 
to disconnect by ambipolar diffusion.
Low mass star formation is very efficient, 
involving $\sim 10^6$ $M_{\odot}$ of 
galactic cold gas at any time, 
in agreement with observed upper limits on cold gas mass. 
We discuss the cooling region surrounding a typical cooling 
site and show that the total X-ray absorption in cold and 
cooling gas is much less that that indicated 
by recent X-ray observations. 
Using a mass dropout 
scheme consistent with X-ray observations and 
dynamical mass to light ratios, we plot 
the global H$\beta$ surface brightness profile 
in NGC 4472 and compare it with 
the smaller contribution from HII gas 
recently ejected from red giant stars.
The lifetime of cooled gas at each cooling site, 
$\sim 10^5$ yrs, is too short to permit dust formation 
and perhaps also gas phase formation of molecules.

\end{abstract}


\section{INTRODUCTION AND OVERVIEW}

Strong X-ray emission from 
luminous elliptical galaxies is clear evidence that the hot
interstellar gas they contain is losing energy. 
Throughout most of the galactic volume, this 
loss of energy does not result in lower temperatures
since the gas is continuously reheated by compression in the 
galactic gravitational potential 
as it slowly moves inward.
In this sense the galactic ``cooling flow'' is a misnomer. 
Ultimately, however, in the central regions 
of the flow the gas density becomes large enough 
for radiative losses to overwhelm dynamical 
compression and the gas cools catastrophically.
For a typical galactic cooling rate, 
$\sim 1$ $M_{\odot}$ per year,
the total amount of gas that cools in a massive elliptical 
over a Hubble time is large, 
several $10^{10}$ $M_{\odot}$, a few 
percent of the total stellar mass.

Remarkably, the amount of cold gas observed in 
ellipticals, either in atomic 
or molecular form, is many orders of magnitude less than 
$10^{10}$ $M_{\odot}$ (Bregman, Hogg \& Roberts 1992).
The mass of central black holes in bright ellipticals is
also relatively small, 
typically less than a few $10^9$ $M_{\odot}$
(Magorrian et al. 1998), 
so the cooled gas cannot be in the holes.
Soft X-ray absorption has been observed 
in some galactic cooling flows, 
indicating masses of cold gas comparable to the 
predicted value, 
but the quantitative significance or reality of this 
absorption is unclear at present. 

In addition to cold gas deposited by cooling flows,
it is possible that additional cold, dusty 
gas is occasionally
delivered to the centers of ellipticals as a result of merging
with gas-rich companion galaxies. 
While this is plausible for some gas-rich ellipticals 
having dusty clouds or lanes, 
if merging were an important source of cold gas 
for all massive ellipticals, 
the merging rate would need to be carefully regulated 
in order to maintain the small amount of cold gas observed 
in normal ellipticals.

For many years the standard theoretical explanation for this
shortage of cooled gas is that it 
has been consumed in forming 
low mass stars (e.g. 
Fabian, Nulsen \& Canizares 1982;
Thomas 1986;
Cowie \& Binney 1988;
Ferland, Fabian, \& Johnstone 1994).
Such young stars must have low masses 
since neither luminous OB stars 
nor Type II supernovae have been observed in normal ellipticals.
Explaining the disappearance of cooled gas 
by invoking the poorly understood physics of star formation 
may seem contrived and the low stellar mass hypothesis 
has led to some ridicule. 

The fate of cooled gas in cooling flows associated with 
clusters of galaxies has received 
most of the observational and theoretical 
attention because of the spectacularly large inferred 
mass deposition rates, ${\dot M} \gta 100$ $M_{\odot}$ yr$^{-1}$ 
(Fabian 1994).
In addition to low mass stars, the apparent soft 
X-ray intrinsic absorption of $N \sim 10^{21}$ 
cm$^{-2}$ indicates that $\sim 10^{10}$ $M_{\odot}$
of cold gas lies within $\sim 100$ kpc of the cluster cores.
Although enormous, this amount of gas would still be only a 
few percent of the total cooled gas based on the estimated 
${\dot M}$, so low mass stars are still the preferred 
endstate for most of the cooled cluster
gas (Allen \& Fabian 1997).
However, as with elliptical galaxies, this amount of cold gas 
should in general be detectable in HI or CO emission 
but has not 
(O'Dea et al. 1994; Ferland, Fabian \& Johnstone 1994;
Vogt \& Donahue 1995; Puy, Grenacher, \& Jetzer 1999), 
amounting to a colossal discrepancy between expectation 
and reality.

In the discussion that follows we revisit the problem of 
cooled gas from the perspective of the galactic 
cooling flow in NGC 4472, 
a large, well-observed elliptical galaxy. 
For relatively nearby ellipticals the threshold for radio 
detection is much lower and the ratio of 
observed to predicted cold gas masses is 
similar to that of more distant cluster cooling flows; 
e.g. H$_2$ and HI are undetected in NGC 4472 with 
an upper limit $10^7$ $M_{\odot}$ 
(Bregman, Roberts \& Giovanelli 1988; 
Braine, Henkel \& Wiklind 1988), 
far below 
the $\sim 10^{10}$ $M_{\odot}$ expected.
Nevertheless, the intrinsic soft X-ray absorbing column 
in NGC 4472, $3 \times
10^{21}$ cm$^{-2}$, and its relatively large covering factor
indicates cool gas masses far in excess of 
the radio upper limit.
The cooling flow in 
NGC 4472 clearly suffers from a miniature version of the same 
cooling flow problems of distant cluster flows. 
But because of its proximity, there is 
a large available body of additional 
observational information for NGC 4472 that 
make it a more appropriate venue to resolve or constrain theoretical 
possibilities for the fate of cooled gas.

But the main advantage afforded by 
large, relatively nearby ellipticals emphasized here is 
that the cooled gas is heated, ionized and therefore {\it illuminated} 
by ultraviolet radiation from highly evolved galactic stars. 
We argue that the diffuse optical line emission from HII gas 
at $T \sim 10^4$ K distributed across 
the central regions of most or all 
bright ellipticals is a direct tracer of cooled gas 
deposited by the hot interstellar gas.
A simple analysis of this HII gas leads to the conclusion 
that hot phase gas is not cooling into a single large cloud 
of neutral gas, but is cooling at a large number 
($\sim 10^6$) of cooling sites located 
throughout the central regions of NGC 4472.
The HII gas provides direct observational support 
for a distributed mass dropout that has been assumed 
by many authors in the past based on their interpretation 
of X-ray surface brightness profiles
(e.g. Thomas 1986; Sarazin \& Ashe 1989). 
As the mass of HII increases at each cooling site, a neutral 
core of very low temperature ($T \approx 10 - 20$ K) 
eventually develops.
We show that these 
neutral cores are very weakly ionized and can undergo gravitational
collapse even in the presence of maximum strength magnetic fields.  
Evidently this collapse results in local star formation.

Another important conclusion from our study of the ionized 
gas in NGC 4472 is that only a very small amount, $\sim 1$ 
$M_{\odot}$, of 
neutral (or molecular) gas can accumulate at each cooling site
before it undergoes gravitational collapse. 
The small mass of collapsing neutral cores is an 
essential requirement for low mass star formation. 
Previous studies (e.g. Ferland, Fabian \& Johnstone 1994) 
have shown that the gas temperature and 
Jeans mass are small deep within 
HI or H$_2$ gas irradiated by X-rays in 
cluster cooling flows, but this does not guarantee that 
massive stars cannot form.
For example, the Jeans mass is 
often very low in Galactic molecular clouds but these clouds 
are also the birthsites for massive OB stars. 
The limited mass of cold gas at each cooling site in NGC 4472 
and other similar ellipticals   
naturally prohibits stars more massive than about 1 $M_{\odot}$
from forming.

The low mass star formation 
process we propose is also very efficient: the total mass  
of cold gas at all cooling sites in NGC 4472  
at any time is very small, consistent 
with observed upper limit of cold gas 
($< 10^7$ $M_{\odot}$).
NGC 4472 is an excellent galaxy for studying this 
unique star formation process since 
very little alien gas or stars have been 
recently accreted into NGC 4472 by a merging process. 
Dusty, and therefore accreted, gas is confined to 
within $r \lta 0.05r_e$ (van Dokkum \& Franx 1995; 
Ferrari et al. 1999)

A final advantage of studying cooled gas 
in ellipticals like NGC 4472 is that the neutral
gas formed in the cores of HII regions with temperatures 
$T \sim 10$ K, only lasts for a time, 
$\lta 10^5$ years, that is too short for 
dust (and possibly many molecules) to form.
Although dust and molecules are 
not required for low mass star formation to proceed, these 
components have complicated previous discussions of 
cooling flows in clusters of galaxies where, we assume, the 
cooling process resembles that in NGC 4472.

One shortcoming of our presentation -- as with 
those of previous 
authors -- is that we cannot reconcile 
the observed soft X-ray absorption 
in NGC 4472 with the small amount of cold gas indicated 
by null observations of HI and CO emission. 
We assume, without much justification, that these 
contradictory observations will be resolved in favor of the 
radio observations and that the soft X-ray absorption can 
be interpreted in another way.

HII gas in elliptical galaxies can also 
arise from stellar mass loss which is ionized by hot central 
stars (planetary nebulae) and galactic UV radiation.
We begin our discussion below with an argument that 
cooling flow dropout, not stellar 
mass loss, is likely to be 
the main contributor to internally-produced HII gas mass 
observed in nearby ellipticals. 
Then we discuss an elementary model for the HII gas at a 
typical interstellar cooling site and infer from this 
a low global filling factor 
for HII gas within the central region of NGC 4472. 
Next we model 
the cooling of hot gas from $\sim 10^7$ to $\sim 10^4$ K 
with a subsonic flow in pressure equilibrium -- 
this flow is useful in estimating the possible contribution 
of cooling gas to the X-ray absorption.
This is followed by a discussion of the temperature, 
ionization level and gravitational instability of neutral cores 
at the centers of HII cooling site clouds.
This part of our presentation follows rather closely 
several previous 
discussions, but serves to illustrate 
that the more spatially concentrated radiative transfer in 
spherical geometry still allows low temperatures and low 
mass star formation within these cores.
We then show from observational considerations 
that magnetic fields play little or no role in inhibiting 
the compression of interstellar gas 
as it cools from $10^7$ to $10^4$ K 
and from theoretical considerations that even the strongest 
observationally allowed magnetic fields are unlikely 
to inhibit the final collapse of neutral cores to stellar 
densities.
At the end of our presentation we discuss the effects of 
galactic gravitational forces and stellar collisions on 
cooling site clouds.
Finally, to stimulate further observations of optical 
emission lines, 
we present a surface brightness map of NGC 4472 
showing all the major components: stars, X-ray emitting gas,
and HII gas from interstellar dropout and stellar ejecta.

\section{NGC 4472: A PROTOTYPICAL ELLIPTICAL GALAXY}

For quantitative estimates in the following discussion, 
we use a specific galaxy, NGC 4472, a massive E2 
galaxy associated with the Virgo cluster.
NGC 4472 has been extensively observed at X-ray frequencies 
with {\it Einstein} HRI 
(Trinchieri, Fabbiano, \& Canizares 1986) and 
with ROSAT HRI and PSPC (Irwin \& Sarazin 1996).
The radial variations of hot gas density and temperature
based on these X-ray data are illustrated in 
Brighenti \& Mathews (1997a).
Although the outer region of the X-ray image of 
NGC 4472 is distorted, 
possibly by ram pressure interaction with ambient Virgo gas, 
the azimuthally averaged radial 
variation of electron density in NGC 4472 is typical of other
bright ellipticals (Mathews \& Brighenti 1998).

The most likely region for low mass star formation in
NGC 4472 is the volume within 
$\sim 0.1r_e$ where $r_e = 8.57$ kpc
is the effective or half-light radius 
at a distance of 17 Mpc.
The gas that cools in NGC 4472 
cannot all collect at the origin, 
nor is it likely that most of the cooling occurs at 
very large galactic radii where the radiative cooling rate
($\propto n^2$) is inefficient.
Brighenti \& Mathews (1999a) have shown that 
if all the cooled gas accumulates 
at or near the very center of the galaxy, $r \lta 100$ pc, 
the remaining uncooled hot gas there 
is locally compressed and becomes very hot, 
but this is not observed. 
Alternatively, if most of the cooling and 
low mass star formation occurs in 
$0.1 \lta r/r_e \lta 1$, then the 
extremely close agreement between the 
total mass inferred from X-ray data 
and the known stellar mass in this region 
would be upset (Brighenti \& Mathews 1997a). 
Finally, there is good evidence from 
observed gas abundance and temperature gradients 
that hot interstellar gas is flowing 
inward within $\sim 3r_e$ through the optically bright 
regions of NGC 4472 
(Brighenti \& Mathews 1999a),
so it is unlikely that a significant number of
low mass stars could form at $r \gta 3 r_e$.
It is most interesting therefore 
that HII optical line emission in 
H$\alpha$ + [NII] lines 
is observed just in the region of NGC 4472 
where low mass star formation is most expected, 
$r \lta 0.24 r_e$ (Macchetto et al. 1996).

\section{SEVERAL SOURCES OF HII GAS}

In addition to interstellar 
gas cooling from the hot phase, cold gas is continuously 
expelled from stars throughout the galaxy 
as a result of normal stellar evolution. 
The total rate that mass is supplied by a 
population of old stars in NGC 4472 
is ${\dot M} = \alpha_*(t_n) M_{*t} \approx 1$ $M_{\odot}$ 
yr$^{-1}$ where $M_{*t} = 7.26 \times 10^{11}$ $M_{\odot}$ 
is the stellar mass in NGC 4472 and 
$\alpha_*(t_n) \approx 1.7 \times 10^{-12}$ yr$^{-1}$ 
is the specific mass loss rate from a single burst of 
star formation after $t_n = 13$ Gyrs (Mathews 1989).
The supply of gas from stars is comparable to the 
rate that gas is observed to cool from the hot phase: 
${\dot M} = \left({2 \mu m_p / 5 k T}\right)
L_{x,bol} \approx 2.5 M_{\odot}$ yr$^{-1}$,
where $m_p$ is the proton mass and 
$L_x \approx 7.2 \times 10^{41}$ ergs s$^{-1}$ is
the bolometric X-ray luminosity of NGC 4472 
at a distance of 17 Mpc.

Several lines of evidence suggest that most of the 
gas lost from stars in ellipticals 
is dissipatively and conductively 
heated and rapidly merges with the general hot interstellar 
cooling flow.
Gas lost from orbiting stars inherits stellar velocities 
which, when dissipated in shocks or by thermal conductivity, 
equilibrates to the virial 
temperature of the stellar system, $T \sim 1$ keV.
However, the
stellar virial temperature is about 30 percent 
lower than that of the more extensive dark halo.
As cooling flow gas slowly flows inward from the halo
into the stellar region, it is cooled by $\sim 0.3$ keV 
as it mixes with slightly cooler virialized  
gas ejected from local stars (Brighenti \& Mathews 1999a).
This produces the positive temperature gradients 
observed within a few $r_e$.
In addition, the iron, silicon and other 
elements supplied by the stars 
increases the metal abundance in the hot interstellar gas   
as it slowly flows toward 
the galactic center within $\sim r_e$, 
producing negative abundance gradients (Matsushita 1997). 
These observations indicate that most or all of the 
gas ejected by stars merges with the hot gas phase. 
For simplicity,
in the following discussion we ignore the HII contribution 
from stellar mass loss, but return to this question 
in \S 11.
This is contrary to the hypothesis of Thomas (1986) that 
gas ejected from stars 
remains largely neutral and collapses into (very) 
low mass stars without joining the hot phase. 

The assimilation of stellar ejecta into the hot 
interstellar gas is greatly accelerated 
by dynamical and thermal processes 
resulting from the orbital motion of mass-losing 
stars through the cooling flow (Mathews 1990). 
Rayleigh-Taylor and other instabilities shred the ejected 
gas into many tiny cloudlets, greatly increasing the surface 
area presented to the hot cooling flow gas and 
their rapid dissipation by conductive heating. 
In addition, neutral clumps of gas expelled from stars 
always have ionized outer layers which are easily 
ablated and reformed; this results in a rapid and complete 
disruptive heating of the clump. 
In contrast, cold gas produced as gas cools directly from the 
hot interstellar phase is necessarily formed in the local 
rest frame of the cooling flow gas so the
violent dynamical instabilities that accompany stellar 
mass loss are not expected. 
After $\sim 10^6$ years, however, 
the denser cooling region may begin to fall 
in the galactic gravitational field (see \S 10), possibly 
leading to some disruption at the cloud  
boundary (Malagoli et al. 1990; Hattori 
\& Habe 1990). 
Assuming that radiative cooling from the 
hot interstellar phase occurs, 
as gas cools through HII temperatures it 
is thermally protected by surrounding gas 
at intermediate temperatures ($10^4 < T < 10^7$ K) where 
the thermal conductivity is very low.
The global kinematics of 
the two HII gas components 
of internal origin are quite different. 
HII regions produced 
by stellar ejecta will initially tend to mimic 
local random and systematic 
stellar motions while HII gas arising from cooling gas 
will initially share the velocity of local hot gas.

A third source of HII gas 
in ellipticals are the ionized parts 
of gas acquired in recent merging events such as 
the small dusty clouds within $\sim 0.05r_e$ of the center
of NGC 4472 (van Dokkum \& Franx 1995). 
This gas is spatially disorganized and is dynamically 
unrelaxed. 
Dust is another clue of its external origin since 
gas formed by cooling from
the hot phase should be nearly dust-free
due to sputtering (Draine \& Salpeter 1979;
Tsai \& Mathews 1995; 1996) and may not have time 
to grow dust in the gas phase (\S 7).

Our interest here is with the HII component produced 
as gas cools from the hot phase 
and we assume that this component dominates 
the optical line emission in NGC 4472.

\section{THE INVERSE HII REGION}

A small cloud of HII gas that has cooled from the 
hot interstellar medium 
is photoionized by stellar UV radiation arriving 
at its outer boundary;
this is the inverse geometry of 
normal HII regions ionized by a central star. 
We suppose that the HII cloud is spherical 
and that the electron density $n_e$ 
and temperature $T = 10^4$ K are uniform throughout 
the HII gas. 
The spatial uniformity of the HII density is essentially 
unaffected by 
small local gravitational fields due to internal 
stars, the neutral core in the cloud if one exists, 
or the HII gas itself.
The mass of these HII clouds located  
at the centers of local cooling sites 
slowly increases with time. 
The first step in understanding the evolution of HII 
clouds is to determine the maximum size 
and mass that can be ionized by stellar UV 
in the central regions of NGC 4472.
This size depends on the HII electron density and 
the mean intensity of galactic UV starlight $J_{uv}(r)$. 

The intensity of ionizing radiation can be determined by 
an appropriate integral over the galactic stellar distribution.
For this we assume a de Vaucouleurs stellar distribution similar 
to that in NGC 4472, with an 
effective radius of $r_e = 8.57$ kpc 
and an outer maximum stellar radius of 25$r_e$. 
The stellar density and mass are given 
to a good approximation by 
$$\rho_* = \rho_o (b^4 r/r_e)^{-0.855} 
\exp[-(b^4 r/r_e)^{1/4}]$$
and
$$M(r) = M_o \gamma(8.58, [b^4 r/r_e]^{1/4})$$
where $b = 7.66925$ and $\gamma(a,z)$ is the 
incomplete gamma function (Mellier \& Mathez 1987).

If the de Vaucouleurs distribution extends to infinity,
the total mass would be
$M_t = M_o \Gamma(8.58) = 1.6582 \times 10^4 M_o$
where $M_o = 16 \pi \rho_o (r_e/b^4)^3$.
It is natural to normalize 
the density coefficient $\rho_o$
to fit the de Vaucouleurs distribution 
for NGC 4472 in the region $0.1 \lta r/r_e \lta
1$ where the X-ray and stellar mass determinations agree, 
i.e. $\rho_o = 3.80 \times 10^{-18}$ gm/cm$^3$.
When the stellar distribution is truncated 
at $25 r_e$ the total mass $6.97 \times 10^7$ $M_{\odot}$ 
is only about 1 percent less than 
an infinite stellar distribution having the same $\rho_o$.

At every radius 
$$x = r/r_e$$ 
in the de Vaucouleurs distribution 
the mean stellar column density $\tilde{J}$ 
can be found by integrating over solid angle,
$$\tilde{J} = {1 \over 2} \int_{-1}^1 \tilde{I}(\mu,x) d \mu
~~~{\rm where}~~~\tilde{I}(\mu,x) = \int \rho_*(x) d \ell.$$
Here $\mu = \cos \theta$ and $\ell$ is the distance back along 
a ray at angle $\theta$ extending through the 
de Vaucouleurs stellar distribution $\rho_*(x)$.
The mean column density $\tilde{J}(x)$ 
(in $M_{\odot}$ of stellar matter cm$^{-2}$ ster$^{-1}$)
is calculated
using the exact solution for the mass intensity along
each ray and averaging over solid angle with 192
Gaussian divisions.
Then $\tilde{J}(x)$ 
is converted into 
an approximate mean intensity of ionizing photons 
${\cal J}(x)$ by using the conversion factor 
$7.3 \times 10^{40}$ ionizing photons per second 
per solar mass of old stars 
(Binette et al. 1994; Brighenti \& Mathews 1997b):
$${\cal J}(x) = 7.3 \times 10^{40} \tilde{J}(x)~~
{\rm cm}^{-2}~{\rm ster}^{-1}~{\rm s}^{-1}.$$
The local density of ionizing photons is then 
$n_{iph}(x)  = 4 \pi {\cal J}(x)/c$.
A plot of $\rho_* (x)$ and ${\cal J}(x)$ in NGC 4472 is shown in 
Figure 1;
the variation of ${\cal J}(x)$ (dashed line) and 
therefore $n_{iph}(x)$ 
is much flatter than the stellar density $\rho_*(x)$ (solid line).

The electron density in HII clouds $n$ can be found 
from 
$$n(x)  = (n T)_{ism}/ T_{HII}$$
where $T_{HII} = 10^4$ K and
$(nT)_{ism}$ is the pressure of the hot ISM that 
reproduces the X-ray observations of NGC 4472.
Plots of $n_{ism}(x)$ and $T_{ism}(x)$ can be found 
in Brighenti \& Mathews (1997b).

The Stromgren condition for the maximum radius $r_s$ of a 
spherical, fully ionized HII cloud is approximately  
\begin{equation}
{\cal L} = n^2 \alpha_B (4/3) \pi r_s^3
\end{equation}
where ${\cal L}$ is the total number of ionizing 
photons incident on (and absorbed by) 
the cloud per second and 
$\alpha_B = 2.6 \times 10^{-13}(T/10^4)^{-0.8}$ 
cm$^3$ s$^{-1}$ is the Case B recombination coefficient.
For an isotropic radiation field, a good approximation near 
the galactic center ${\cal L} = 4 \pi r_s^2 {\cal F}$, 
where ${\cal F} = \pi {\cal J}$ is the flux of ionizing 
photons incident on the surface of the HII cloud. 
This is equivalent to ${\cal L} = (1/4) 
n_{iph} c 4 \pi r_s^2$ which is reasonable.
The Stromgren radius is then
$$r_s = {3 \pi {\cal J}(x) \over n^2 \alpha_B}.$$
The density of HII gas is 
$\rho = n M f_{\rho}$ where $f_{\rho} = 5 \mu /(2 + \mu) = 1.20$, 
assuming $\mu = 0.61$ for the molecular weight.
The total mass of the Stromgren sphere is 
$$m_s = {4 \over 3} \pi r_s^3 n m_p f_{\rho} 
= {36 \pi^4 {\cal J}^3 m_p f_{\rho}
\over n^5 \alpha_B^3}$$
where $m_p$ is the proton mass. 
Some imprecision is expected since we have ignored 
those ionizing photons that pass 
through the HII cloud unabsorbed. 
However, calculations of the transfer of 
ionizing radiation in the inverse HII region indicate 
that Equation (1) is accurate to $\lta 5$ percent.

Figure 2 illustrates the radial variation of 
electron density $n_e = (\rho/m_p) (2 + \mu)/5\mu$ 
in HII gas (solid line) with galactic radius 
in NGC 4472 and the corresponding local inverse
Stromgren radius $r_s$ (long dashed line).
Within the radius where H$\alpha$ is observed
in NGC 4472, $x = r/r_e \lta 0.24$ we find  
$r_s \approx 0.3 - 0.8$ pc, $n_e \approx 20 - 90$ cm$^{-3}$,
and the mass of a typical Stromgren cloud is
$m_s \approx 2$ $M_{\odot}$. 
The radial column density in an HII Stromgren cloud is
typically $N_s = n_e r_s \approx 1.2 \times 10^{20}$
cm$^{-2}$.

Hot gas is assumed to be cooling at numerous sites 
throughout this central region of NGC 4472 
and the cooling is made visible by optical line emission 
from the HII regions.
The mass of any
particular HII cloud increases slowly with time, 
supplied by local cooling from the hot gas phase
or by dissipative merging of clouds.
Presumably, new HII clouds are continuously forming 
from the cooling interstellar gas at 
newly-formed cooling sites and 
old sites and associated clouds are disappearing.
But we suppose that the mean age of cooling sites is long 
compared to 
the time required for typical HII clouds 
to reach the Stromgren mass;
in this case the average cloud can be approximated 
with Stromgren parameters.
Notice also that $r_s \ll r$ so that even the largest 
HII clouds are 
very small compared to their distance to the center of the galaxy.


\section{GEOMETRY OF HII AND COLD GAS}

We propose that most of the extended Balmer line emission 
in ellipticals arises from a multitude of 
HII clouds at or near their Stromgren radii.
If so, the total volume within clouds occupies only a 
tiny fraction $f_F$ of the galactic volume 
within the H$\alpha$-emitting 
region of NGC 4472, $r \lta 2$ kpc.
The filling factor $f_F$ 
can be estimated by comparing the total 
volume of HII required to produce the observed Balmer line
luminosity to the apparent volume from which optical
line emission is observed.

In most optical observations, such as those of 
Macchetto et al (1996), H$\alpha$ (6562 \AA) 
is blended with two nearby [NII] lines at 6584 and 6548 \AA. 
The total flux observed by Macchetto et al. (1996) in all 
three lines is 
$F_{lines} = 17.30 \times 10^{-14}$ ergs cm$^{-2}$ 
s$^{-1}$.
Observations at higher resolution reveal that the 
F([NII] 6584)/F(H$\alpha) \approx 1.38$ and 
F(6584)/F(6548) = 3.0.
Combining all these ratios, 
and adopting Case B conditions $F_{H\alpha}/F_{H\beta}
= 2.86$, the H$\beta$ flux 
from NGC 4472 is 
$F_{H\beta} = 2.13 \times 10^{-14}$ ergs cm$^{-2}$ s$^{-1}$ 
and the total luminosity from all HII emission is
$L_{H\beta} = 4 \pi D^2 F_{H\beta} = 7.34 \times 10^{38}$ 
erg s$^{-1}$,
assuming a distance of $D = 17$ Mpc to Virgo.

How many dust-free Stromgren clouds are required to produce this
total luminosity?
The H$\beta$ luminosity of a single Stromgren cloud is 
$$\ell_{\beta,s} = n_e^2 \epsilon_{\beta} (4 \pi/3) r_s^3
= 1.5 \times 10^{31} n_e^2 r_{spc}^3 ~~ {\rm ergs}~{\rm s}^{-1}$$
where $\epsilon_{\beta} =
1.0 \times 10^{-25}$ erg cm$^3$ s$^{-1}$
is the H$\beta$ emissivity at $T = 10^4$ K.
For typical values of $n_e$ and $r_{spc}$ (in parsecs)  
in the central galaxy $x \lta 0.25$,
$\ell_{\beta} \approx 3 - 7.5 \times 10^{33}$ ergs 
s$^{-1}$.
Therefore, about ${\cal N}_{cl} = 10^5 - 10^6$ Stromgren
clouds are required to account for 
the Balmer line luminosity observed.
The HII filling factor is found by comparing 
the volume of all HII gas 
$V_{cl} = L_{H\beta} / 
\langle n_e \rangle^2 \epsilon_{H\beta} = 
2.5 \times 10^{60}$ cm$^3$
(assuming $\langle n_e \rangle = 50$ cm$^{-3}$)
with the total volume of the H$\beta$-emitting region,
$V_{tot} = (4/3)\pi (0.24 r_e)^3 = 1.1 \times 10^{66}$ 
cm$^{-3}$.
The filling factor of HII gas 
$f_F = 2 \times 10^{-6}$ is very small, 
consistent with our proposition that HII emission arises
from many small clouds and with earlier 
estimates of $f_F$ (Baum 1992).
If $\sim 1$ $M_{\odot}$ of hot gas 
cools each year in NGC 4472, 
then the mass of each cloud will grow quite slowly, 
$\sim 10^{-6} - 10^{-5}$ $M_{\odot}$ yr$^{-1}$, 
requiring $t_s \sim 2 \times 10^5 - 2 \times 10^6$ years
to form a typical Stromgren cloud.
The total mass of all the HII emitting gas in NGC 4472 is
$M_{II} = \langle n \rangle M f_F V_{cl} \approx 1.2 \times 10^5$
$M_{\odot}$, similar to values in the literature
but here evaluated using a consistent physical model.

A small filling factor also implies 
that each HII cloud is exposed to the 
unabsorbed stellar UV emission from the entire galaxy, 
provided the cloud system is approximately spherical.
The ``optical depth'' for intersecting a Stromgren cloud 
across the optical line-emitting region within 
$r_t = 0.24 r_e$ is 
$\tau = \pi r_s^2 r_t {\cal N}_{cl}/V_{tot} \approx 0.006
- 0.06$.
Since $\tau \ll 1$ the clouds do not shadow each other.
In reality $\tau$ could be larger 
(i) if the typical cloud crossection is much less 
than $\pi r_s^2$ ($\tau \propto r_s^{-1}$) 
or (ii) if the cloud 
system were not spherical; a disk-like 
configuration could result from galactic rotation.
In any case, 
the assumption that each HII cloud is exposed to the 
full, unabsorbed stellar UV emission 
is likely to be a reasonably good approximation.

Combining previous results, the average column depth 
that HII gas presents to X-radiation throughout the 
galactic core, $N \sim N_s \tau 
\sim 10^{18} - 10^{19}$ cm$^{-2}$, is much less than the value 
$N \sim 3 \times 10^{21}$ cm$^{-2}$ that best fits the 
observed X-ray continuum (Buote 1999). 
The size that an HII cloud presents to absorbing
X-rays is larger than $r_s$ since we have
ignored the extended cooling region around each
cloud with temperatures between $10^6$ and $10^4$ K 
in which X-rays can still be absorbed.
This assumption will be justified below. 

The total mass of HI or H$_2$ gas observed in the 
central regions of NGC 4472, $M_{cold} < 10^7$ $M_{\odot}$, 
is another potential source of X-ray absorption.  
If this mass of cold gas were arranged in a disk of thickness 
of the X-ray absorbing column $N = 3 \times 10^{21}$ cm$^{-2}$, 
located in the galactic core and oriented with its 
symmetry axis along  
the line of sight, it would have a radius $< 370$ pc, 
somewhat larger than the faint patch 
of dust observed by van Dokkum \& Franx (1996).
However a cloud of size 
370 pc obscures only $\sim 0.007$ of the total 
X-ray luminosity of NGC 4472 and would 
therefore produce negligible X-ray absorption.
The true X-ray absorption is very probably much 
less than $3 \times 10^{21}$ cm$^{-2}$. 
The observation of NGC 4472 by Buote using 
the $\sim 4$' beam of ASCA also included the nearby gas-rich 
dwarf irregular galaxy UGC 7636 
(Irwin \& Sarazin 1996;
Irwin, Frayer \& Sarazin 1997) which 
may be the source of the X-ray absorbing column attributed 
to NGC 4472 if its covering factor is 
sufficiently large.

Although it seems likely that interstellar 
magnetic fields are 
important in the centers of ellipticals
(Mathews \& Brighenti 1997; Godon, Soker \& White 1998), 
it is remarkable that the 
observed optical line emission does not indicate strong 
fields in the HII gas.
Typical HII densities in bright ellipticals determined from 
[SII] 6716/6731 line ratios are 
$\sim 100 - 200$ cm$^{-3}$ (Heckman et al. 1989; 
Donahue \& Voit 1997), 
similar to (or even a bit larger than)
the values found here for NGC 4472 (Figure 2).
(Unfortunately, we have been unable to find a determination 
of the HII density specific to NGC 4472.)
This suggests that the HII gas density is not being diluted 
by magnetic support, i.e. $B^2/8 \pi < 2 n k T$
or $B < 70\mu$G in the HII gas.
HII densities of $\sim 100$ are 
also supported by comparing the ionization parameter 
$U = n_{iph}/n_e$ for pressure equilibrium 
HII gas in NGC 4472 (short dashed line in Figure 2)
with values that characterize the entire 
observed line spectrum.
Within $\sim r_e$ in NGC 4472 
$\log U \approx -3.3$, very similar to 
values of $U$ required to reproduce LINER type 
spectrum typically observed in ellipticals
(e.g. Johnstone \& Fabian 1988); this provides an 
independent check on our HII gas density and 
${\cal J}$ near the center of NGC 4472.

The apparent absence of magnetic support in the HII gas 
is interesting since the hot phase gas is required 
to have fields of at least several $\mu$G 
at large galactic radii to explain  
Faraday depolarization of radio sources and distant quasars
(Garrington \& Conway 1991).
Interstellar fields $\gta 1 \mu$G 
can be generated in a natural way by 
stellar seed fields and 
turbulent dynamo action in the hot gas 
(Mathews \& Brighenti 1997).
As the gas density increases by $\sim 1000$ 
when it cools  
from the hot phase to HII temperatures, 
a field of 1 $\mu$G would grow to 100 $\mu$G 
if flux is conserved, $B \propto \rho^{2/3}$.
The initial field in 
the hot gas in $r \lta 0.24 r_e$ would need to be 
surprisingly small, $\lta 0.7\mu$G, 
to evolve into the rather small fields allowed in HII 
clouds, $B \lta 70\mu$G, implied by typical 
electron densities. 
Small HII fields 
can be understood if local cooling sites form in 
interstellar regions 
having lower than average fields; 
in pressure balance, lower fields require 
higher hot gas densities which cool preferentially.
Alternatively, it is possible that field reconnection 
has been very efficient during cooling, 
implying a disorganized field 
and considerable stirring motion during the cooling process.

\section{COOLING SITE GAS DYNAMICS}

Cooling sites in the hot interstellar 
gas are initiated in regions of low entropy 
(i.e. low temperature, high density) 
which cool preferentially by radiative losses.
Entropy fluctuations can be generated by a  
variety of complex events:
stellar mass loss, occasional Type Ia supernovae, 
sporadic mergers with other nearby (dwarf) galaxies, 
and differential 
SNII heating events and outflows that occurred in 
pregalactic condensations.
Due to the complicated nature of these 
interactions, it is difficult to predict 
the amplitudes and mass scales of the entropy inhomogeneities 
so the detailed nature of the
initial cooling process remains unclear.
However, once cooling commences, the gas flow toward 
the cooling site may evolve toward a simple profile provided 
entropy fluctuations in the hot gas are not 
too severe over the flow region. 

We now describe such 
a model for cooling site flow
into an individual HII cloud in which 
the flow is largely subsonic and little influenced
by magnetic stresses.
In the subsonic
limit we can assume that the flow is isobaric.
The thermal energy equation is
$$\rho { d \varepsilon \over dt}
= {P \over \rho} { d \rho \over dt}
- {\rho^2 f \Lambda \over m_p^2}$$
where $P = (k/\mu m_p) \rho T$
is the constant pressure of the ambient hot gas,
$\varepsilon = 3P/2\rho$ is the specific thermal energy,
and $\Lambda(T)$ is the Raymond-Cox cooling coefficient
which must be accompanied by the factor
$f = (2 + \mu) (4 - 3\mu)/(25 \mu^2) = 0.58$.
The corresponding temperature evolution is
$${dT \over dt} = - {2 \over 5}
{\mu^2 P f \over k^2} { \Lambda \over T}.$$
If the converging flow at a cooling site is
regular over times $10^6 - 10^7$ years, then
steady state is a good approximation,
${\dot m} = -\rho u 4 \pi r^2$, and
$$ {d r \over d t} =
- { {\dot m} k \over 4 \pi \mu m_p P}
{ T \over r^2}$$
describes the velocity of a fluid element.
The spatial variation of gas temperature in the cooling
site flow is found by dividing the equations above:
$${dT \over dr} =
{8 \pi \over 5}
{ \mu^3 P^2  m_p f \over k^3 {\dot m} }
{ \Lambda r^2 \over T^2} 
\equiv K \Lambda(T) { r^2 \over T^2} .$$
We solve this equation by integrating outward from
the boundary of a Stromgren cloud of typical
radius $r_s = 0.3$ pc and
temperature $T = 10^4$ K assuming
a uniform pressure typical of the inner galaxy,
$P = 2 \times 10^{-10}$ dynes.
Once $T(r)$ is known, the density and flow velocity
can be found from the remaining equations.
Cooling site
flows for ${\dot m} = 10^{-6}$ and $10^{-5}$
$M_{\odot}$ yr$^{-1}$ are illustrated in Figure 3. 
The flow is subsonic at every radius.
When $\Lambda(T)$ is slowly varying, the 
temperature varies approximately linearly with radius, 
$T \propto (K \Lambda)^{1/3} r$.
These simple solutions indicate that the
cooling region in which $T \lta 10^6$ K is
at most about twice the Stromgren radius
so the extended cooling region
can only increase the soft X-ray crossection of each
cloud by $\lta 4$, not enough to account for the
indicated soft X-ray absorption.

We do not consider 
conductive heating and thermal disruption 
of cooling site flows  
by ambient hot gas since this would be incompatible 
with our basic assumption that the hot interstellar 
gas is cooling over a large galactic volume. 
However, once a cooling site is
formed, it cannot be easily
reversed by thermal conductivity.
The HII cloud that forms at the center of a cooling
site is surrounded by
a protective sheath of gas at intermediate temperatures
in which the thermal conductivity is extremely low
($\kappa \propto T^{5/2}$) even in the 
absence of magnetic fields.
By comparison, gas expelled from evolving red giant stars
is directly exposed to interstellar gas at $T \sim 10^7$ K,
greatly enhancing the likelihood of conductive
dissipation of that gas. 
Finally, thermal conductivity is much less 
important in galactic cooling flows 
than in hotter cluster 
cooling flows.

The soft X-ray emission from the cooling region shown
in Figure 3 is enhanced by the larger density but
suppressed by the lower mean temperature in the flow.
For the flows in Figure 3
the X-ray intensity in the ROSAT band (0.2 - 2.5 keV)
at $r = 0$ (ignoring absorption in the HII cloud) is
increased by $\sim 10$ compared to the same emission
from non-cooling interstellar gas if it occupied 
the same volume as the flow.
However, since the volume of the cooling site flow
is tiny compared to the entire galaxy, the mean X-ray
intensity incident on the HI core is not sensibly
increased by emission from the surrounding 
cooling site flow.
Nor is external X-radiation incident on the 
HII cloud appreciably absorbed 
within the cooling site flow.

\section{MASS OF NEUTRAL CORES}

As an HII cloud grows in mass at a cooling site,
its radius must eventually exceed the Stromgren value, 
causing a neutral or molecular core to develop at its center.
As the cloud and core mass increase further, 
the core will become unstable to gravitational collapse 
if the gas there is sufficiently cold.
We shall determine the core mass 
at which collapse first occurs and thus determine 
an upper limit to the mass of stars that can form.
For simplicity we assume that the core is primarily atomic 
hydrogen. 

Several authors have discussed the possibility and 
implications of gas-phase dust formation 
and growth in cooled gas 
in cluster cooling flows (Daines, Fabian \& Thomas 1994;
Fabian et al. 1994; Voit \& Donahue 1995).
However, dust may not have time 
to form or grow in NGC 4472 
because of the insufficient available time. 
For a lower limit to the dust formation time consider 
assembling graphite grains with carbon atoms that 
stick with every collision and ignoring carbon atom 
depletion from the gas phase.
Grains of radius $a_g = 10^{-5}$ cm that can absorb optical 
light have mass $m_g = 4 \pi a_g^3 \rho_c/3$ where 
$\rho_c = 2$ gm cm$^{-3}$ is the density of graphite.
The rate that a grain mass grows is 
${\dot m}_g = n_c {\bar v}_c \pi a_g^2 m_c$ where 
$n_c = 4 \times 10^{-4} n_{HI}$ is the density of C atoms 
that move at mean velocity 
${\bar v}_c \approx (8 k T / \pi 12 m_p)^{1/2}$. 
For $n_H = 10^5$ cm$^{-3}$ and $T = 10$ K 
(see below) the grain 
formation time is 
$t_g = 4 \rho_c a_g / 12 m_p n_c {\bar v}_c \approx 2 \times 10^5$
years. 
Since this minimum value for 
$t_g$ is comparable to the lifetime 
$t_s \sim 10^5 - 10^6$ yrs 
of neutral or molecular cores (see \S 8 below), 
it is unlikely that much dust can form.
In addition, 
molecules cannot form in gas phase interactions 
without grains (Herbst 1987).
Formation of H$_2$ via the H$^-$ ion is also problematic 
in galactic cooling flows because of the high probability of 
H$^-$ photodetachment by galactic starlight. 
We shall not consider these details here.

An essential assumption in estimating the mass of 
neutral cores is that the accumulation of gas in  
cooling site clouds is slow compared to the free fall time
in the neutral core.
If neutral gas accumulates faster than it can collapse under 
self-gravity, then more massive and luminous stars 
can form, contrary to observation.
Since $\gta 10^5 - 10^6$ HII clouds
are collectively processing the entire cooling rate of the
galaxy, about $10^{-6} - 10^{-5}$ $M_{\odot}$ 
flows into each cloud per year.
Neutral cores of density $n_c$ 
are in pressure equilibrium with 
the surrounding HII and ambient hot interstellar gas, 
$n_c T_c \approx (nT)_{ism}$, ignoring factors of 
order unity involving molecular weights. 
Near the center of NGC 4472 the hot gas 
pressure is $(nT)_{ism} \approx 10^6$ 
K cm$^{-3}$ so the gas density in the neutral cores,
$$n_c \approx (nT)_{ism}/T_c \approx 10^5 
~(T_c /10~{\rm K})^{-1}~~~{\rm cm}^{-3},$$ is large, 
similar to star-forming cores in Galactic molecular clouds.
If the cores become gravitationally unstable, they collapse
in $t_{dy} \sim (3 \pi/32 n_c m_p G)^{1/2} \approx 10^5$ years.
Since the typical cloud takes 
a longer time $\sim 3 \times 10^5 - 3 \times 10^6$
to form, 
the required slowness seems to be guaranteed, but not
by a wide margin. 

As the masses of molecular cores slowly increase, they
must eventually become gravitationally unstable and form 
into stars.
The condition for this to occur can be estimated 
using the virial theorem with surface pressure,
\begin{equation}
2 {3 \over 2} {kT_c \over \mu_n m_p} M_{HI}
= {3 \over 5} { G M_{HI}^2 \over r_c}
+ 3 P {4 \over 3} \pi r_c^3
\end{equation}
where $P  \approx 2 \times 10^{-10}$ dynes is the ambient 
interstellar pressure near the center of NGC 4472.
Although the temperature and 
density within the cold core are expected to 
develop radial gradients due to gravity forces,
in this simple virial model 
the temperature and density within the HI core are 
uniform, but the core 
pressure is enhanced by self-gravitation.
Magnetic forces and rotation are ignored.

Consider a family of equilibrium models for the HI core 
in which the core mass $M_{HI}$ and external pressure $P$ 
are held constant but the core radius $r_c$ 
and internal temperature $T_c$ are allowed to vary.
For very large cores, the gravity term in the virial 
equation is small and $r_c \propto T_c^{1/3}$;
for very small cores, the surface pressure term is 
negligible and $r_c \propto T_c^{-1}$.
Therefore a minimum 
temperature $T_m$ exists where $dT_c/dr_c = 0$.
If the true temperature of an HI core is less than $T_m$,
no equilibrium model is possible and the core must collapse.

It is interesting therefore to examine the locus of 
critically stable models with $T_c = T_m$ for a wide variety 
of core masses, holding the external pressure fixed.
Differentiating the virial equation and setting $dT_c/dr_c = 0$ 
provides an expression for the core radius 
when $T_c = T_m$:
$$r_m = \left( {G M_{HI}^2 \over 20 \pi P } \right)^{1/4}.$$
For each $T_c > T_m$ there are two radii that satisfy 
virial equilibrium, but only the larger of these 
roots is stable and physically relevant.
For this solution, $r_c \propto T_c^{1/3}$ holds at large 
$T_c$, but as $T_c \rightarrow T_m$ the radius decreases to $r_m$ 
where gravitational collapse commences.
For a given core mass and external pressure,
the density and temperature of these marginally stable 
configurations are 
\begin{equation}
\rho_m = {3 M_{HI} \over 4 \pi} 
\left( {G M_{HI}^2 \over 20 \pi P} \right)^{-3/4}
\end{equation}
and 
\begin{equation}
T_m =
\left({\mu_n m_p \over k}\right) \kappa
P^{1/4} G^{3/4} M_{HI}^{1/2}
\end{equation}
where
$$\kappa = 
\left[ { (20 \pi)^{1/4} \over 5}
+ {4 \pi \over 3 (20 \pi)^{3/4}}\right] = 0.751.$$
Of most interest is 
the critically stable core mass 
for any temperature $T_m$:
\begin{equation}
M_{HI,m} = { (k T_m/\mu_n m_p)^2 \over \kappa^2 
P^{1/2} G^{3/2} } $$
$$= 1.56  
\left( {T_m \over 10 {\rm K}} \right)^2
\left( {P \over 2 \times 10^{-10} {\rm dy}} \right)^{-1/2}
~~M_{\odot}.
\end{equation}
If neutral
or molecular cores can cool to $\lta 10$ K, 
collapse will occur and all stars
that form must be less massive than $\sim 1.56$ 
$M_{\odot}$ -- 
perhaps very much less depending on  
subsequent mass fragmentation. 
Since the mass of each HII cloud increases slowly 
compared to the free fall time of the neutral core, 
the core mass will not increase further as it collapses.
Low mass star formation is inevitable. 

The HI number density is related to the core density 
by $n = (\rho/m_p)(4 - \mu_n)/(3 \mu_n)$ where 
$\mu_n = 1.27$ is the HI molecular weight.
For critically stable clouds,
the neutral column density across the core radius 
depends only on the external pressure,
$$N_m  =  n_m r_m = {3 \over 4 \pi m_p}
{4 - \mu_n \over 3 \mu_n}
\left( {20 \pi P \over G} \right)^{1/2}$$
$$~~~~=  1.1 \times 10^{23} 
\left( {P \over 2 \times 10^{-10}{\rm dy}} \right)^{1/2}
~~{\rm cm}^{-2}\eqno{(6)}$$
\addtocounter{equation}{1}
and is independent of the core mass $M_{HI}$.
Finally, the pressure in marginally unstable HI cores 
is $P_m = 4P$ for all $M_{HI}$; this core overpressure 
relative to that in the ambient 
hot interstellar gas is due to 
the additional gravitational compression unique to this
region.

Figure 4 illustrates the mass and radii of the HII and HI 
components of marginally stable clouds as a function 
of core temperature $T_m$ 
with external pressure 
$P = 2 \times 10^{-10}$ dynes similar to that in 
the inner regions of NGC 4472, $r\lta 0.1r_e$.
The minimum stable core mass varies inversely with molecular 
weight, $M_{HI,m} \propto \mu_n^{-2}$.
If hydrogen is molecular, $\mu_{H2} = 2.33$ and the leading 
coefficient in Equation (5) decreases from 1.56 to 0.44.
Masses for 
critically stable cloud configurations having molecular cores 
are shown in Figure 4 with heavy lines.
Also shown in Figure 4 is the number density inside 
critically stable HI cores.

Equation (4) for the minimum temperature for stable 
core configurations 
also holds at every radius within the core, 
even if the core density is not constant with radius, provided 
we consider a series of homologous cores of constant mass 
in which all radii scale
linearly and gas density scales as $\propto r^{-3}$.

\section{TEMPERATURE AND STABILITY OF NEUTRAL CORES}

The gas temperature within the neutral cores 
is determined by a 
balance of heating by absorption of X-rays or starlight 
and cooling by far infrared line emission.
Thermal X-rays are produced 
by hot cooling flow gas throughout 
the galaxy. 
Infrared line emission results from 
thermal excitation of low-lying fine structure levels.
If the mean temperature in the 
core $\langle T \rangle$ based on these physical processes 
is less than the minimum temperature for gravitational 
equilibrium $T_m$, 
then gravitational collapse and star formation will occur.
In estimating the temperature in HI cores, approximations 
will be used that tend to maximize $\langle T \rangle$; 
this will lead to an overestimate of the critically 
unstable core mass that collapses. 
The procedure we use is similar to that in other recent 
papers 
(O'Dea et al. 1994; Ferland, Fabian \& Johnstone 1994;
Vogt \& Donahue 1995), but our treatment of the 
radiative transfer is 
spherical and we include the possibility of UV heating 
and absorption of X-rays in the HII cloud envelope.

\subsection{Photoionization and Heating by Stellar UV}

Absorption of stellar UV radiation by those elements 
having ionization potentials less than 1 Rydberg 
-- C, Fe, Mg, Si, etc. -- can 
be important in HI heating and ionization
in some circumstances. 
To estimate this possible energy source in HI cores,
we consider only the most abundant
contributor to HI heating, carbon ionization.
Carbon can be ionized by radiation in a 
narrow radiation window, 11.26 -- 13.6 eV, 
beyond which radiation is absorbed in 
the surrounding HII region.

The importance of C $\rightarrow$ C$^+$ 
as a source of heating and ionization in HI cores can be 
estimated by balancing ionization and recombination 
rates, just as in an inverse HII region. 
For an HI core of total radius $r_{HI}$ we expect 
that carbon will be largely ionized in 
an outer region $r_{C+} < r < r_{HI}$ 
where $r_{C+}$ is determined by 
$${\cal L}_c \approx n_e n_{C+} \alpha_c(T) 
(4/3) \pi (r_{HI}^3 - r_{C+}^3).$$
Here 
$n_e \approx n_{C+} \approx n_C \approx A_c n_{H}$
is the number density of (fully) ionized carbon 
and $A_c = 3.62 \times 10^{-4}$ is the 
solar carbon abundance by number relative to hydrogen,
The C$^+$ + e$^-$ $\rightarrow$ C recombination coefficient is 
$\alpha_c = 1.4 \times 10^{-10} ~ T^{-0.607}$ 
cm$^3$ s$^{-1}$ (Hollenbach \& McKee 1979; 1989).
The equation above can be 
solved for the radius of the carbon ionization front,
$$r_{C+} = r_{HI} (1 - \gamma)^{1/3}$$
where
$$\gamma(r_{HI}, T) = 
{ 3 \pi {\cal J}_c \over r_{HI} (A_c n_H)^2 \alpha_c }.$$
If $\gamma > 1$ then $r_{C+} = 0$ and 
carbon is ionized throughout the HI core.

The mean photon intensity of radiation in the carbon 
ionizing window at any galactic radius 
${\cal J}_c(r)$ 
(photons cm$^{-2}$ s$^{-1}$ ster$^{-1}$) can 
be estimated from  
the observed flux at a wavelength 
at the center of the C ionization window, 
$ \lambda_c \approx 1100$ \AA.
For NGC 4472 this flux is 
$F_c = 2.0 \times 10^{-12}$ ergs cm$^{-2}$ s$^{-1}$
(Brown, Ferguson, \& Davidsen 1995; 1996).
At a distance of 17 Mpc the luminosity in the 
carbon window is 
$L_c = 6.9 \times 10^{40}$ ergs s$^{-1}$ or 
$9.5 \times 10^{28}$ ergs s$^{-1}$ per solar mass 
in stars.
This last ratio can be used to convert the mean 
stellar mass surface density 
$\tilde{J}(r)$
($M_{\odot}$ cm$^{-2}$ ster$^{-1}$) to the mean 
intensity of carbon window radiation 
at any radius in NGC 4472, 
$J_c(r) = h \nu_c {\cal J}_c(r)$
$= 7.3 \times 10^{40} \tilde{J}(r)$
$= 1.30 \times 10^{-12} {\cal J}(r)$.
Using values appropriate to the central regions of 
NGC 4472, $r \lta 0.1r_e$, 
\begin{equation}
\gamma 
= 0.025 \left( {T \over 10 {\rm K} } \right)^{2.607}
\left( { r_{HI} \over 0.03 {\rm pc}} \right)^{-1}
\end{equation}
where $P = 2 \times 10^{-10}$ dynes is assumed.
We show below that the fraction of the HI cores 
in which carbon is ionized is very small, 
$\gamma \ll 1$, so heating by stellar UV is unimportant.

\subsection{Photoionization and Heating by Thermal 
X-rays}

The bolometric 
X-ray energy density $u_x(r)$ and mean intensity 
$J_x(r)$ can be found 
at every radius in the galaxy by averaging the 
intensity of hot gas thermal 
radiation over solid angle.
The intensity from each direction 
is found by integrating
the equation of transfer over the 
X-ray emissivity $\epsilon_x(n_e,T,z)$ which depends
on the local hot gas 
electron density, temperature and abundance.
The bolometric X-ray emissivity for solar abundances 
can be defined as in Sutherland \& Dopita (1993),
$$\epsilon_x = {n_p^2 \over 4 \pi} \Lambda_{sd}(T,z) 
= \left( {4 - 3\mu \over 2 + \mu} \right)^2 
{n_e^2 \over 4 \pi} \Lambda_{sd}(T,z) $$
(ergs cm$^{-3}$ s$^{-1}$ ster$^{-1}$) 
where $n_p$ is the proton density 
and $\mu = 0.61$ is the molecular weight of fully ionized gas.
To evaluate the emissivity $\epsilon_x(r)$ we use 
the radial variation of hot gas density
$n_e(r)$ and temperature $T(r)$
for NGC 4472 from Brighenti \& Mathews (1997a),  
assuming solar abundance.
The mean intensity $J_x(r)$ is found by averaging the 
X-ray intensity over solid angle 
with 192 Gaussian divisions. 
The mean bolometric X-ray energy density 
$u_x(r) = 4 \pi J_x(r)/c$ in ergs cm$^{-3}$ 
is plotted in Figure 1.
In the central regions of the galaxy ($r \lta 0.1 r_e$) 
we find 
$u_{x} \approx 10^{-14}$ 
ergs cm$^{-3}$ so that $u_x \sim 0.1 u_{iph}$.

The specific X-ray photon intensity $\phi(E)$
(photons cm$^{-2}$ s$^{-1}$ keV$^{-1}$) 
incident on the cloud is determined 
with the MEKAL option of the XSPEC software code.
We assume that the X-ray spectrum in NGC 4472 can be adequately 
characterized by an isothermal 1 keV plasma with solar abundances.
The mean intensity at photon energy $E$ incident 
on the cloud located at galactic radius $r$ 
can be normalized to the bolometric energy density 
$u_x(r)$ found earlier: 
$$J(E,r) = {c u_x(r) \over 4 \pi}
{ E \phi(E) \over \int E \phi(E) dE }
~~~{ {\rm ergs} \over 
{\rm cm}^{2} {\rm s} {\rm ster}
{\rm keV} }.$$

The intensity incident on the outer HII boundary of a 
cloud is assumed to be isotropic, $I(E) = J(E)$.
The mean intensity at the HII-ISM 
boundary is slightly 
lower than the ambient $J(E)$ due to absorption within 
the cloud.
In computing the X-ray transfer into the cloud,
the incident intensity from 0.01 to 10 keV is divided into 
30 logarithmically-spaced energy bands.
Only a limited energy range is relevant. 
Photons of energy $\lta 0.03$ keV cannot penetrate the 
HII gas surrounding HI cores while photons of energy 
$\gta 4$ keV pass through critically unstable neutral cores 
without absorption.
As X-radiation penetrates the cloud it is absorbed 
with a crossection $\sigma(E)$ 
modeled after that of Morrison \& McCammon (1983).
The neutral cores are divided into twenty equally spaced 
computational zones $r_k$
and an additional zone is assigned to the HII gas.
The mean X-ray intensity $J(E,r)$ and flux $F(E,r)$ is 
determined at every zone boundary and at the center of 
the cloud for each of the 30 energy bands 
of mean energy $E_m$.
The equation of transfer within the cloud
is solved to find the mean intensity $J(E_m,r_k)$ and
flux $F(E_m,r_k)$ using the same
192-order Gaussian divisions routine described earlier, 
but now including absorption.
The mean photon intensity $J(E_m,r_k)$ is shown 
in Figure 5 at three radii 
-- $r_{HII} = 0.206$ pc, $r_{HI} = 0.041$ pc, 
and $r = 0$ -- in an HII-HI cloud 
having core mass $M_{HI} = 3.5$ $M_{\odot}$.
For this particular neutral core only X-rays  
with energies $0.1 \lta E \lta 4$ keV are absorbed in 
the core.

The total X-ray heating rate within the radial zone between 
$r_{k}$ and $r_{k+1}$ is 
\begin{equation}
H_k = \varepsilon(x)
{| F_{k+1} 4 \pi r_{k+1}^2 - F_{k} 4 \pi r_{k}^2 |
\over (4 \pi /3) (r_{k+1}^3 - r_k^3)}~~~
{ {\rm ergs} \over {\rm sec}~{\rm cm}^{3}}
\end{equation}
where $F_k = \Sigma F(E_m,r_k) \Delta E_m$ is the 
integrated flux at $r_k$.
For simplicity we assume that the energy 
of Auger electrons is equal to the photoionization threshold 
for each element.
Only a fraction $\varepsilon(x)$ of the total X-ray 
energy absorbed is available to heat the gas.
Each photoionization by a photon of energy 
$E$ produces an energetic primary electron 
with energy $E - \chi$ where $\chi \ll E$ is the 
ionization potential.
This primary electron subsequently ionizes 
many additional
H atoms, creating on average about $E/\Delta \epsilon$ secondary 
electrons where $\Delta \epsilon = 0.037$ keV 
(Shull \& van Steenberg 1985; Glassgold, Najita, \& Igea 1997).
Collisional excitation of atomic levels of H and He 
by these secondary electrons produces 
line radiation, 
most of which escapes and does not heat the gas.
Only when the electron energy drops below 
the excitation threshold for L$\alpha$ (10.2 eV)
will all its energy be converted to heat,
usually by Coulomb exchanges among free electrons.
The heating efficiency $\varepsilon(x)$ depends on the 
ionization level $x = n(H^+)/n(H)$ which is found 
by balancing rates for recombination and ionization:
$$x^2 n(H)^2 \alpha_B(T) = (1 - x)n(H) \zeta$$
where $n(H)$ is the total number density of both neutral 
and ionized hydrogen and $\alpha_B(T) = 2.91 \times 10^{-11}
[T(r)/10 {\rm K}]^{-0.602}$ cm$^3$ s$^{-1}$ is the 
low temperature 
Case B recombination coefficient (Ferland 1993).
The ionization rate per neutral hydrogen is 
$$\zeta(r) = \sum_m {4 \pi J(E_m,r) \Delta E_m 
\over 1.60 \times 10^{-9} E_m} \sigma(E_m) 
\left[ {E_m \over \Delta \epsilon} \right] ~~~{\rm s}^{-1}$$
where $1.60 \times 10^{-9}$ converts keV to ergs.
The heating efficiency in Equation (8) 
$$\varepsilon(x) = {\rm max}
\{ 0.10, ~0.9971[ 1 - (1 - x^{0.2663})^{1.3163} ] \}$$ 
is an approximate fit to the Monte Carlo results 
of Shull \& van Steenberg (1985) but with a constant 
value at very low $x$ 
(Xu \& McCray 1991; Voit \& Donahue 1995).

\subsection{Cooling of HI Cores by Line Emission}

Most of the cooling in the HI gas is due to excitation of 
fine structure levels: 370 and 610 $\mu$m emission from CI
and 63 and 145 $\mu$m emission from OI
(Hollenbach \& McKee 1979, 1989; 
Ferland, Fabian, \& Johnstone 1994).
In estimating the cooling rate of the HI gas we 
determine the exact occupation density of the 
three lowest fine structure levels of CI and OI, 
taking HI collisional rate coefficients and $A$ values 
from Hollenbach \& McKee (1979; 1989).
By considering only these dominant cooling levels, 
we underestimate 
the cooling rate 
which conservatively overestimates the HI temperature 
and the mass that 
becomes gravitationally unstable.

The principal cooling lines of CI and OI 
become optically thick within the HI core.
We allow for this by determining the mean escape probability 
${\bar \epsilon}$ for each line.
For each pair of levels $(u,l)$ the line 
center optical depth across the core radius 
for a Doppler broadened line is 
$$\tau_{o,lu} = 1.74 \times 10^{-18} N
\lambda_{ul}^3 {\cal A}_{el} A_{ul} {g_u \over g_l} 
\left({ A_e \over \langle T \rangle } \right)^{1/2}$$
where $N$ is the column density in the HI core, 
$\lambda_{ul}$ is the wavelength of the transition 
in $\mu$m, 
$A_{ul}$ is the Einstein emission rate, 
${\cal A}_{el}$ and $A_e$ are the cosmic abundance 
and atomic number of C or O,
$g_u$ and $g_l$ the degeneracies of the upper and lower levels,
and $\langle T \rangle$ is the mass-weighted mean temperature 
over the HI core.
Typically $\tau \sim 10 - 100$ for strong cooling lines.
The mean escape probability ${\bar \epsilon}$ 
for each line is found 
by averaging the mean escape probability from a 
uniform sphere
$$ \epsilon(\tau) = {3 \over 4 \tau}
\left[ 1 - {1 \over 2 \tau^2} 
+\left( {1 \over \tau} + {1 \over 2 \tau^2} \right)
e^{-2\tau} \right]$$
over the Doppler width of the line,
$${\bar \epsilon}(\tau_o) = {2 \over \pi^{1/2}}
\int_0^\infty dx e^{-x^2} \epsilon(\tau_o e^{-x^2}).$$
Here $\tau_o$ is the optical depth at line 
center, $x = (\nu - \nu_o)/\Delta\nu_D$ is a dimensionless 
frequency and $\Delta \nu_D = (\nu_o / c) (2 k 
\langle T \rangle /A_e m_p)^{1/2}$ is the Doppler width.
To an excellent approximation in the range 
$10^{-1} \lta \tau_o \lta 10^6$
the mean escape probability is
$$ {\bar \epsilon}(\tau_o) = {1 \over 1 + f(\tau_o)}
~~~{\rm where}~~~ f(\tau_o) = { 0.6 \tau_o 
\over 1 + 0.04 \tau_o^{0.26} },$$
which allows ${\bar \epsilon}(\tau_o)$
to approach unity as $\tau_o \rightarrow 0$. 
The mean escape probability for cooling line emission 
is only an approximate substitute for a proper radiative 
transfer calculation including temperature gradients. 
Unfortunately, the escape of line emission can 
be enhanced by turbulent motion which 
cannot be easily estimated.

\subsection{Results: Gravitationally Unstable 
Cloud Cores}

As the mass of the HI core $M_{HI}$ is varied, 
a family of critically unstable HII-HI cloud models 
can be generated, all with identical 
ambient parameters similar to those in 
the optical line emitting region in NGC 4472: 
external pressure $P = 2 \times 10^{-10}$ dynes, 
bolometric X-ray energy density 
$u_x = 1.7 \times 10^{-14}$ ergs s$^{-1}$ and 
stellar UV ${\cal J}_{uv} = 1 \times 10^8$ 
photons cm$^{-2}$ s$^{-1}$ ster$^{-1}$. 
The HII temperature is fixed at $10^4$ K 
and the radiative transfer is solved ignoring 
gravity.
The temperature and density in each HI core model 
is initially set to $T_m$ and $\rho_m$ throughout. 
As new temperatures $T(r_k)$ are found from the 
energy balance at each radius,
the density and radius 
of each HI zone is adjusted 
to maintain uniform pressure $4P$ 
throughout the core, while preserving each zone mass.
The X-ray radiative transfer is repeated 
with the new zone spacing, producing new temperatures 
for the next iteration.
The core structure converges after 3 or 4 iterations. 

Figure 6 illustrates the variation of $n(r)$, $T(r)$, $M(r)$, 
$N(r)$, and $x(r)$ within a neutral core of mass 
3.5 $M_{\odot}$. 
This configuration is just globally unstable to gravitational 
collapse:
the mass-weighted mean temperature for this core is 
$\langle T \rangle = 14.6$ K while the 
minimum temperature for stable models having this core mass 
is $T_m = 15.0$ K.
Table 1 lists core and HII properties of 
several cloud models having different HI core masses. 
For each cloud the global gravitational stability 
is determined by comparing the 
mean temperature $\langle T \rangle$
with the minimum stable temperature $T_m$ within 
radius $r_{HI}$.
It is clear from Equation (7) that 
$\gamma \ll 1$ 
for relevant values of $r_{HI}$ and $\langle T \rangle$ in 
Table 1, i.e. heating and ionization by 
X-rays dominate UV photoelectric 
processes throughout almost all of the HI core volume. 

The two temperatures listed in Table 1, $\langle T \rangle$
and $T_m$, refer to global averages over the entire HI core.
However, these temperatures can be evaluated at 
any interior radius in the core, as discussed earlier.
When this is done, we find that 
$\langle T \rangle / T_m$ has a broad minimum within 
the HI structure.
Figure 7 illustrates the internal 
variation of $\langle T \rangle / T_m$ 
within HI core mass for cores of five total masses.
It is seen that HI cores are completely 
stable as long as the total mass 
does not exceed $\sim 2.5$ $M_{\odot}$.
However, when $M_{HI} \approx 3$ $M_{\odot}$ 
the entire region between about 1.2 and 2.8 $M_{\odot}$
becomes unstable.
Evidently collapse first begins 
at an interior zone of mass $\sim 2$ $M_{\odot}$ when the 
total core mass is $2.5 \lta M_{HI} \lta 3.0$ $M_{\odot}$.
Therefore, 
as a cloud grows in mass at at rate 
$\sim 10^{-6}$ $M_{\odot}$ yr$^{-1}$, stars of mass 
$\lta 2$ $M_{\odot}$ would collapse at the center of 
a typical cloud every $\sim 2 \times 10^6$ years. 
If a single star formed it would be of type $\sim$F0V 
on the main sequence. 
It is also possible that collapsing cores 
fragment into a distribution of stellar masses described 
by some IMF.
For example, if the IMF is Salpeter between 0.03 and 2 
$M_{\odot}$, then the maximum stellar mass that actually 
forms is about $0.95$ $M_{\odot}$. 
The brighter members of 
such a population of stars would be optically luminous. 

The results in Table 1 confirm that HI core temperatures 
are indeed as low as the normalization used 
in Equation (5).
Because of the spherical concentration of X-ray energy 
in our calculation, 
the HI gas temperatures in Table 1 are larger than 
those discussed by Ferland, Fabian \& Johnstone (1994), 
but their conclusions about low mass star formation 
still hold.
The maximum mass of stars that can form at cooling sites 
is $\lta 2$ $M_{\odot}$.
But this instability threshold is clearly overestimated.
The unstable core mass would be significantly 
smaller, for example, 
if some of the core gas were molecular, 
if more radiative cooling transitions were considered,
or if a small amount of internal 
turbulence lowered the optical depths of 
cooling line radiation.

Is massive star formation also excluded at larger 
galactic radii? 
To explore this we performed a series of 
calculations like those in Table 1 but using external
parameters expected at $r = r_e = 8.57$ kpc in NGC 4472:
pressure $P = 10^{-11}$ dynes, 
X-ray energy density $u_x = 6.1 \times 10^{-16}$ 
ergs cm$^{-3}$ and UV intensity 
${\cal J}_{uv} = 4 \times 10^6$
photons cm$^{-2}$ s$^{-1}$ ster$^{-1}$.
With these parameters we find that the lowest 
gravitationally unstable HI core mass 
is 4.5 $M_{\odot}$ 
where $\langle T \rangle \approx T_m \approx 9$ K.
The lower pressure at $r = r_e$ 
generates somewhat larger critically unstable core masses, 
even though the mean HI temperature is lower (Equation 5).
Fragmentation 
into a Salpeter IMF (from 0.03 to 4.5 $M_{\odot}$) 
during collapse 
would limit the typical maximum stellar mass to  
$\sim 2$ $M_{\odot}$ (i.e. A5V to F0V).
Such stars, having highly radial orbits, could 
contribute to the H$\beta$ absorption line
indices observed in elliptical galaxies
(Gonzalez 1993; Worthey 1994;
Faber et al. 1995; Worthey, Trager, 
\& Faber 1995)
which have been interpreted as evidence for 
a population of young 
stars of intermediate mass in ellipticals.
The observed H$\beta$ absorption index usually strengthens 
toward the centers of ellipticals (in $r \lta 0.5r_e$), 
indicating lower mean stellar ages;  
this is consistent with 
cooling flow dropout which should 
also be centrally enhanced. 
The more prominent H$\beta$ absorption seen 
in ellipticals of lower luminosity may 
reflect the influence of lower interstellar pressure 
in Equation (5) and correspondingly larger unstable 
core masses.

\section{COMPLICATIONS: MAGNETIC FIELDS AND ROTATION}

For a definitive understanding
of star formation in elliptical galaxies or elsewhere, 
it is necessary to consider the influence 
of magnetic fields and rotation, 
both of which are intensified during  
the two major compressions that lead to gravitational 
instability and star formation in elliptical galaxies.

During the initial cooling condensation from 
$T = 10^7$ to $10^4$ K, neither rotation nor magnetic 
energy density appear to have strongly restricted the collapse.
Any fragmentation due to 
rotation during this initial cooling phase
probably serves only to increase the multitude of local 
cooling sites, each containing a small HII region.
Remarkably, the frozen-in 
amplification of magnetic energy during this 
initial compression has not occurred since 
optical observations indicate that the 
HII gas pressure equals the pressure of the 
local hot interstellar gas and is therefore undiluted by 
additional magnetic support.
The second cooling compression 
by another factor of $\sim 10^3$ occurs 
as the core gas recombines and the temperature 
falls from $10^4$ K to $\sim 10$ K. 
During this compression, 
magnetic fields will intensify if they 
remain attached to the weakly ionized 
plasma and rotation should also increase. 
Rotation during this final stage may 
lead to subsequent fragmentation into 
even smaller masses, although the entire cloud is 
optically thick in the main cooling lines.
A helpful 
interplay between magnetic breaking and angular momentum 
redistribution can be expected as stellar densities are approached. 
In this section we discuss a limited aspect of this 
general problem, the loss of magnetic stresses 
by reconnection and ambipolar diffusion 
as neutral gas recombines and compresses to typical 
HI core densities, $\sim 10^5$ cm$^{-3}$. 

Although difficult to prove rigorously, reconnection losses 
could in principle greatly reduce magnetic stresses 
and their inhibiting influence on compression. 
We have already noted that the observed HII gas pressure 
in bright ellipticals 
restricts fields in the HII gas to $B \lta 70$ $\mu$G, 
suggesting that some reconnection losses 
occurred during compression 
from local interstellar to HII densities. 
Reconnection could be stimulated by 
turbulence and disordered fields expected in the 
hot interstellar gas 
(Mathews \& Brighenti 1997) which, when compacted to HII 
densities, result in locally anti-directed fields.
In the second compression phase,  
as HII gas condenses into HI cores, 
the density increases by another factor of 1000 
and the final flux-frozen fields would then be 
$\lta 7000$ $\mu$G.
Fields this large in the HI core would 
strongly inhibit gravitational instability and star formation 
and increase the minimum core mass necessary 
for gravitational collapse.
However, it is possible that reconnection losses 
also dominate this second cooling phase if the 
field is sufficiently disordered 
(Friaca \& Jafelice 1998).
If reconnection is wonderfully efficient, 
the field-free estimates we have made here 
for the maximum unstable stellar mass do not 
require further modification.

Fortunately, physical conditions in HI cores allow for 
efficient slippage of the magnetic field relative to the 
neutral gas.
For densities and field strengths of interest here, 
the charged-neutral collision rate is much less than the 
Larmor frequency for both ions and electrons,
so ambipolar diffusion should prevail over 
Ohmic dissipation or Hall effect (Wardle \& Ng 1999). 
As the field moves relative to the (essentially stationary) 
neutral component in HI cores, ion-neutral drag must balance 
magnetic tension,
$\langle \sigma v \rangle n_i \rho v_d 
\approx {B^2 / 4 \pi r}.$
Here $\sigma$ is the momentum transfer cross-section.
When $v_d$ is less than some transition velocity 
$v_{tr}$, charge-neutral interactions 
are dominated by the polarizibility 
$\alpha$ of atomic hydrogen, 
$\sigma \approx 2.4 \pi (e^2 \alpha/ \mu_m v)^{1/2}$ and 
the rate coefficient $\langle \sigma v \rangle \approx 
2 \times 10^{-9}$ cm$^3$ s$^{-1}$ is independent of temperature
(Osterbrock 1961; Nakano \& Umebayashi 1986). 
When $v_d \gta v_{tr}$ the crossection is limited by the geometrical
size of the particles involved, $\sigma \approx \pi a_B^2$, where
the Bohr radius $a_B$ is appropriate for 
H$^+$ or e$^-$ collisions with H (Mouschovias \& Paleologou 1981).
When applied to the equation above, the drift velocity $v_d$ is 
either linear or quadratic.
For collisions in weakly ionized hydrogen, the transition velocity 
$v_{tr} \approx 230$ km s$^{-1}$ is found by equating the two 
crossections.

The time required for a field to leak through a weakly ionized 
neutral core $t_{ad} \approx r/v_d$ in the linear regime is 
$$t_{ad} \approx { 
4 \pi r^2 \langle \sigma v \rangle x n^2 \mu_{HI} m_p
\over B^2 }$$
where $x = n_e/n$ is the degree of ionization.

To illustrate the efficiency of ambipolar diffusion in
collapsing HI 
cores we can exploit the result that larger fields diffuse out 
faster, i.e. $t_{ad} \propto B^{-2}$.
We begin by considering two simple limits:
(1) initially uncompressed neutral gas that 
is supported by magnetic stresses and (2) dense neutral gas  
after it has collapsed to core densities 
approaching gravitational instability.
For the initial state, consider a gas which has become 
largely neutral, but is supported by a field 
of maximum strength, $B \sim 70\mu$G, 
so that its density $n \approx 100$ cm$^{-3}$ is still that of 
the HII gas.
As long as $\beta^{-1} \equiv (B^2/8\pi)/P \lta 0.5$ 
the column density in 
such a core of mass $\sim 1$ $M_{\odot}$ still
resembles that of typical HI cores in Figure 6, 
so $x \sim 10^{-6}$.
Under these circumstances the ambipolar diffusion time is 
$$t_{ad}~^{(1)}  \approx 
6.7 \times 10^{3} 
\left( { r \over 10^{18} {\rm cm} } \right)^2
\left( { n \over 100 {\rm cm}^{-3} } \right)^2 $$
$$\times~~\left( { x \over 10^{-6} } \right)
\left( { B \over 50 \mu{\rm G} } \right)^{-2}~~~{\rm yrs}.$$
This time is much less than the free fall time 
for field-free cores, 
$t_{ff} \approx 5 \times 10^6$ yrs,
so the strongest possible initial fields cannot arrest 
the initial compression toward typical HI core densities. 

After the HI gas compresses to the final core density 
but just before gravitational collapse occurs, 
then $n \approx 10^5$ cm$^{-3}$ 
and a magnetic field $\sim 100\mu$G 
is comparable with the gas pressure.
In this final state before core collapse, 
the ambipolar diffusion time 
$$t_{ad}~^{(2)}  \approx 
5 \times 10^{6}
\left( { r \over 10^{17} {\rm cm} } \right)^2
\left( { n \over 10^5 {\rm cm}^{-3} } \right)^2 $$
$$\times~~\left( { x \over 3 \times 10^{-7} } \right)
\left( { B \over 100 \mu{\rm G} } \right)^{-2}~~~{\rm yrs}$$
is now longer than the core freefall time,
$\sim 10^5$ yrs or the dynamical time 
$\sim 10^6$ yrs desired to process cold gas rapidly into stars, 
involving observationally small HI and CO masses. 
Some additional field losses by reconnection may be required.

However, this estimate of the field diffusion time in the final core
fails to acknowledge that 
field diffusion occurs steadily during the 
collapse to core densities.
Suppose for example that the field varies as if it were frozen-in,
$B \propto n^{2/3}$, then the ratio of field diffusion to 
free fall time at each density during the collapse varies as 
$t_{ad}/t_{ff} \propto n^{1/2} x$ which 
increases only modestly as $n$ increases by $\sim 1000$, 
holding $x$ approximately constant. 
(We use the linear drift regime, which generally applies.)
If this time scale ratio is normalized 
to the initial time ratio above,
$t_{ad}~^{(1)}/t_{ff} = 0.0013$, then 
$t_{ad}~^{(2)}/t_{ff} = 0.04$ when 
core densities are reached, which remains much less than unity. 
In the absence of a detailed calculation, this crude estimate 
suggests that magnetic fields can slip away from HI cores 
rapidly compared to the dynamical times involved 
in the gravitational evolution of HI cores.

When field diffusion is important, however, 
the field is spatially redistributed, not destroyed. 
In a rigorous calculation it would be necessary to 
follow the subsequent evolution of the field which 
may accumulate in the outer regions of the core. 
If left behind after star formation, 
such magnetically enhanced 
regions may ultimately become buoyant in the galactic 
field and merge again with the HII gas.
In addition, the thermal balance in the neutral gas will 
be altered by heating due to field diffusion and 
reconnection, but this heat may be carried away by efficient 
radiative losses.
In summary, it appears that magnetic stresses may not be 
an overwhelming problem in restricting neutral cores from 
gravitational collapse.
How magnetic and rotation effects influence the subsequent 
history of the star after collapse ensues will not be 
entered into here.

\section{DYNAMICAL INFLUENCES ON COOLING SITE CLOUDS}

Because of their high density relative to the
hot interstellar gas,
HI-HII clouds will tend to fall in the
gravitational field of the galaxy past the hot gas.
As an HI-HII cloud falls toward the center of the galaxy,
the HII envelope should experience a drag force,
so some motion of the HI core away from the center
of the HII envelope could be expected.
For a conservative estimate of this,
suppose that the HI core is
accelerated toward the center of the galaxy
by the galactic gravitational force
but the galactic position of the HII envelope is held fixed.
At a
galactic radius of $r = 0.03 r_e$, the gravitational
acceleration toward the galactic center is
$g \approx G M_*(r)/r^2 \approx 2 \times 10^{-6}$  cm/s$^2$
where $M_* \approx 7 \times 10^9$
$M_{\odot}$ at $r = 0.03 r_e$.
(If the cooling flow is rotating, this acceleration will
be an overestimate.)
The time required for the HI core to move a distance
equal to its own diameter ($2r_c \approx 0.08$ pc),
$t_d = ( 4 r_c / g)^{1/2} \approx 1.5 \times 10^4$ yrs,
exceeds the recombination time
in the HII gas, $t_{rec} = 1/(n \alpha_B)
\approx 1.4 \times 10^3$ years.
Even in this extreme example which maximizes the
differential motion of HII and HI,
gas can exchange between the HI and HII phases rapidly
enough to maintain the
neutral core near the center of the cloud.

How many galactic stars are momentarily passing
through HII clouds or their cold cores at any time?
From Figure 1 the stellar density at $r = 0.1 r_e$
is $\rho_* \approx 2 \times 10^{-22}$ gm cm$^{-3}$ or
$3$ $M_{\odot}$ pc$^{-3}$.
If the average stellar mass is 0.3 $M_{\odot}$,
the stellar density is $\sim 10$ stars pc$^{-3}$.
An HI-HII
cloud with radius of $r_{HII} \approx 0.21$ pc (Table 1)
has volume $\sim 1$ pc$^{-3}$ and should contain
about ten small stars. 
The time-averaged density of stars
within the HII region is comparable to that of the
HII gas itself ($n_H \sim 2$ $M_{\odot}$ pc$^{-3}$).
Moving at
a typical orbital velocity, $\sim 300$ km s$^{-1}$,
a star should pass through the tiny HI core
of radius $r_{HI} = 0.04$ pc only
very infrequently, every $\sim 10^5$ years,
causing little or no disruption.
Finally, we note that the current Type Ia supernova rate 
required to maintain the observed 
interstellar metallicity, 
$\sim 0.03 $ SNe per $10^{10} L_B$ 
every 100 years, is so low that the galactic volume 
occupied by supernova remnants is negligible.
Supernovae are not expected to greatly disturb  
cooling site clouds.

\section{MASS DROPOUT AND EMISSION LINE SURFACE BRIGHTNESS PROFILES}

Optical line emission in ellipticals 
identifies where hot interstellar gas is cooling and 
forming into low mass stars. 
In this section we estimate the H$\beta$ surface brightness 
in NGC 4472 due to 
emission from cooled hot interstellar gas. 
We use a dropout procedure that is 
self-consistent, but not necessarily unique.

One of the least understood but essential components 
in evolutionary models for 
the hot interstellar gas is the cooling dropout mass profile. 
The notion of distributed cooling mass dropout was first 
invoked in steady state cooling flow calculations 
to reduce the central X-ray surface brightness peak 
to better fit the observations 
(Thomas 1986; Thomas et al. 1986; Sarazin \& Ashe 1989;
Bertin \& Toniazzo 1996).
In our recent galactic cooling flow models we 
compare the effect of distributed mass dropout 
on both the apparent (observed) temperature and electron density 
distributions.
Dropout reduces the apparent temperature and increases the 
apparent density.
However, we find that it is the 
dropout contribution to the 
mass to light ratio  
(dynamically determined from stellar velocities) 
that most strongly constrains cooling dropout 
mass profiles (Brighenti \& Mathews 1999b).
One of the simplest dropout assumptions 
that adequately satisfies these constraints 
is the recipe employed by Sarazin \& Ashe (1989);  
we use it here to illustrate the relationship between  
optical line emission and mass dropout.

Following Sarazin \& Ashe (1989), we suppose that the hot gas cools 
at a rate 
${\dot \rho} = q \rho / t_{do}$
that depends on the local hot gas density and the 
cooling time at constant pressure, 
$t_{do} = 5 m_p kT / 2 \mu \Lambda \rho$.  
Here $T$ is the hot phase temperature and 
$q$ is a constant factor that characterizes 
the cooling rate.
Assuming that the $M/L_B$ of low mass stars is 
very large, we find that $q = 1$ satisfies 
reasonably well most of the observational constraints 
for NGC 4472 (Brighenti \& Mathews 1999b).
We suppose that most of the cooling gas is processed through 
mature cooling sites where the typical 
HII cloud mass is comparable to 
the local Stromgren mass $m_s$. 
At each cooling site 
mass is processed into low mass stars 
at a rate $\sim m_s/t_s$ where $t_s \sim 10^5$ yrs, based on 
the filling factor argument.
By continuity, the total rate that 
cooling mass passes through an element 
of galactic volume $dV$ is ${\dot \rho} dV \approx 
d{\cal N} (m_s/t_s)$
where $d{\cal N}$ is the number of $\sim$Stromgren clouds in $dV$.
Each Stromgren cloud has an H$\beta$ 
luminosity $\ell_{\beta,s} \approx 
n \epsilon_{\beta} m_s 
/ \mu m_p$ where
$n = (n T)_{ism}/T_{HII}$ is the electron density of gas 
at $T_{HII} = 10^4$ K 
and $\epsilon_{\beta}$ is the Case B H$\beta$ emissivity.
The mean global H$\beta$ emissivity 
at radius $r$ due to dropout cooling is therefore
\begin{eqnarray}
\varepsilon_{\beta} & = & {d L_{\beta} \over 4 \pi r^2 dr } 
 =  {d{\cal N} \over dV} \ell_{\beta,s} 
= {\dot \rho} {t_s \over m_s}
{n \epsilon_{\beta} m_s \over \mu m_p}\nonumber\\
& = & q t_s 
{ 2 \mu^2 \epsilon_{\beta} \over 5 k T_{HII} }
n_{ism}^3 \Lambda(T_{ism}) 
~~~{ {\rm erg} \over {\rm cm}^{3}~{\rm s} }.
\end{eqnarray}
Notice that $\varepsilon_{\beta}$ is independent of 
the mean cloud mass $m_s$ 
but depends on the time $t_s$ for low mass stars to form 
at each cooling site. 
The H$\beta$ surface brightness $\Sigma_{\beta}(R)$ for this 
dropout model is found by integrating $\varepsilon_{\beta}$ 
along the line of sight through the galaxy at projected 
radius $R$.

For comparison, we also estimate the H$\beta$ surface brightness 
distribution from gas recently ejected from evolving 
stars and ionized by galactic ultraviolet radiation. 
Ejected envelopes from red giant stars 
are not sufficiently massive to produce clouds 
of Stromgren mass. 
The ejection process is complex and slow, 
lasting $\sim 10^5$ years. 
In addition, mass-losing stars orbit rapidly 
through the hot gas.  
This results in violent 
Rayleigh-Taylor disruption in $\sim 10^4$ years 
as the ejected gas
is decelerated and shocked by the hot interstellar gas. 
The huge surface area that 
the highly fragmented stellar ejecta
presents to the hot gas promotes the conductive 
assimilation of this HII gas 
into the local hot gas on timescale $t_{*cond} \sim 10^4$ yrs
(Mathews 1990).
The assimilation time $t_{*cond}$ depends 
on the density of the local interstellar gas, the mean velocity
of mass-losing stars, and other parameters.  
In estimating the H$\beta$ emissivity from stellar ejecta we 
assume that these complications can be subsumed into 
the single parameter $t_{*cond}$ which we regard 
as spatially uniform. 
The rate of supply of mass from evolving stars is proportional 
to the local stellar density, 
${\dot \rho_*} = \alpha_* \rho_*$, where
$\alpha_* = 5.4 \times 10^{-20}$ sec$^{-1}$ is 
appropriate for a Salpeter type IMF.
The local mean 
H$\beta$ emissivity due to stellar ejecta is therefore 
\begin{equation}
\varepsilon_{\beta,*} = {d L_{\beta,*} \over 4 \pi r^2 dr }
= {\dot \rho_*} t_{*cond} {n \epsilon_{\beta} \over \mu m_p}~~~
{ {\rm erg} \over {\rm cm}^{3}~{\rm s} }
\end{equation}
where $n = (n T)_{ism}/T_{HII}$
since pressure equilibrium is reached on very short timescales.

In Figure 8 we illustrate the X-ray surface brightness 
$\Sigma_{ROSAT}$ of NGC 4472 
in the ROSAT energy band (0.2 - 2.0 keV) computed from the 
combined 
density and temperature distributions $n(r)$ and $T(r)$ 
determined from 
{\it Einstein}, ROSAT, and ASCA data 
(see Brighenti \& Mathews 1997a for details). 
Using these same $n(r)$ and $T(r)$ profiles, we have determined 
the H$\beta$ emissivity from Equation (9) and plot the 
corresponding surface brightness $\Sigma_{H\beta}$(dropout) in 
Figure 8, assuming a Sarazin-Ashe dropout coefficient 
$q = 1$.
$\Sigma_{H\beta}$(dropout) is 
normalized so that the total H$\beta$ luminosity within 
$R \approx 0.25 r_e$ matches the observed value 
$L_{H \beta} = 7.34 \times 10^{38}$ erg s$^{-1}$ 
from this region; 
with this normalization the total H$\beta$ luminosity from 
the entire galaxy is somewhat larger: 
$L_{H \beta} = 11.8 \times 10^{38}$ erg s$^{-1}$.
To achieve this normalization 
it is gratifying that the characteristic star formation
time in Equation (9) 
must be $t_s = 2.2 \times 10^5/q$ years, 
very close to the value discussed earlier when $q = 1$. 
This value of $t_s$ normalized to $L_{H \beta}(r < 0.25r_e)$ 
is a consequence of our assumption that most 
H$\beta$ line emission comes from dropout,
any additional HII contribution from stellar mass loss would 
tend to increase $t_s$.

We also plot in Figure 8 
the stellar surface brightness 
$\Sigma_*$ for NGC 4472 using the standard 
de Vaucouleurs relation 
$$\Sigma_*(R/r_e) =  {L_B \over 7.21 \pi r_e^2} 
\exp \{ -7.67 [ (R/r_e)^{1/4} - 1 ] \}$$
which agrees very well with the CCD photometry of 
NGC 4472 by Peletier et al. (1990).
The H$\beta$ surface 
brightness profile $\Sigma_{H\beta,*}$ from gas recently 
ejected from stars using Equation (10) 
is also shown in Figure 8 with two 
assimilation times, $t_{*,cond} = 10^4$ (preferred) 
and $10^5$ yrs. 
These estimates for $\Sigma_{H\beta,*}$, although very 
uncertain, are less than $\Sigma_{H\beta}$(dropout), 
indicating that most of the H$\beta$ observed in 
NGC 4472 is related to mass dropout not stellar mass loss.
This conclusion is not inconsistent with the observation 
of a few bright, young planetary nebulae in ellipticals 
ionized by central stars
(e.g. Arnaboldi, Freeman \& Gerhard 1998).
A considerable amount of observational data 
is available 
on the radial distribution and kinematics
of optical line emission in normal
ellipticals 
(Trinchieri \& di Serego Alighieri 1991; 
Macchetto et al. 1996; Zeilinger et al. 1996;
Pizzella et al. 1997; Trinchieri, Noris \& 
di Serego Alighieri 1997; 
Goudfrooij \& Trinchieri 1998;
Corsini et al. 1999), 
but additional observations of 
optical emission line spectra and surface brightness 
distributions would be useful in understanding the 
physics of mass dropout from the hot interstellar gas.

\section{FINAL REMARKS AND CONCLUSIONS}

Models for the evolution of hot interstellar gas in 
ellipticals that are successful in most other respects 
-- such as explaining the observed radial distribution of 
hot gas density, temperature and abundance -- 
cannot be fully accepted unless they can also account for 
the final disposition of cooled gas. 
Recent evolutionary models for the hot gas in 
NGC 4472 require that a mass 
$M_{cool} \sim 4 \times 10^{10}$ $M_{\odot}$ 
must cool in some fashion over the Hubble time
(Brighenti \& Mathews 1999b).
There are two serious problems in discharging this cold gas 
from the cooling flow: 
(1) Although $M_{cool}$ is much less 
than the total stellar mass $M_{*} = 7.26 \times 10^{11}$
$M_{\odot}$, the mass of cooled 
gas must be deposited within the (inner) 
galaxy in a manner 
consistent with the known stellar mass to light ratio,
and 
(2) The cooling process and the conversion
of cooled gas into low mass stars or some optically
dark equivalent must be supported with 
a physically plausible model 
-- this is the problem we address here.

Our objective in this paper is to further our 
understanding of mass dropout by exploiting 
features unique to elliptical galaxy cooling flows  
that are generally unavailable 
or less obvious in cluster cooling flows.
Foremost among the advantages in galactic flows is that 
the the dropout is illuminated by stellar 
UV ionizing radiation, appearing as HII emission 
visible throughout the inner galaxy. 
By this means we assert that the cooling 
must be widespread, distributed among many cooling sites. 
Moreover, the cooling rate at each site is sufficiently 
slow that HII clouds there can develop neutral 
cores that undergo gravitational collapse 
after only $\sim 2$ $M_{\odot}$ of neutral gas 
has accumulated. 
Only low mass stars can form. 
The conversion of hot gas into low mass stars 
in galactic cooling flows is very efficient, 
with star-formation timescales so short,
$\sim 10^5$ years, that dust and probably molecules 
cannot form. 
Globally, 
this cooling dropout model 
involves very little mass of neutral gas at any time, 
${\cal N}_{clouds} m_s \sim 10^6$ $M_{\odot}$ and satisfies 
the low upper limits on observed 
HI + H$_2$ gas in NGC 4472 ($< 10^7$ $M_{\odot}$). 

In a similar investigation 
several years ago, Ferland, Fabian \& Johnstone (1994) 
studied the thermal structure of cold clouds directly 
illuminated by X-rays from cluster cooling flow gas.
Specifically,
they solved the plane parallel transfer 
of $\sim 5$ keV thermal radiation 
into a cold, non-magnetic, 
dust-free slab, including a large number of coolants and 
chemical reactions. 
They showed that deep within the irradiated slab, 
$N \gta 10^{20}$ cm$^{-2}$, the gas 
becomes molecular and the temperature 
falls below 10 K where the Jeans mass is only 
a few tenths of a solar mass, implying that 
low mass stars are likely to form. 

Our study here is a complementary 
extension of this idea, including several 
additional considerations. 
We consider fewer atomic and chemical processes 
than the 
Ferland, Fabian \& Johnston (1994) calculation, 
but many complicated processes involving dust 
and molecules may be irrelevant 
in galactic cooling flows due to the 
short lifetime of the cooled gas before 
collapsing to stellar densities.  
We have shown that the more concentrated spherical 
radiative transfer of X-rays into  cooling sites 
with star-forming 
neutral cores does not increase the core gas temperature 
appreciably. 
Unlike the plane parallel calculation, where 
Jeans unstable regions could in principle 
merge to form  
more massive stars, in our spherical model 
the {\it total} mass of gas 
in the cold cores prior to gravitational collapse -- 
$\lta 2$ $M_{\odot}$ -- absolutely precludes 
the formation of massive stars 
which are never observed in normal ellipticals. 

The simple model proposed here is unlikely 
to apply to every elliptical galaxy. 
Unlike massive ellipticals like NGC 4472,
the stellar systems in 
low luminosity ellipticals ($M_V \lta -21$) 
differ in many important details: 
they are rotationally supported, 
have more nearly isotropic stellar velocities, 
have dense power-law stellar cusps, are disky, etc.
(Faber et al. 1997).
Brighenti \& Mathews (1997b) have shown that 
large, rotationally supported cold gaseous 
disks can form in these low luminosity ellipticals 
and that normal, luminous star formation 
in the disk may account for the systematically 
higher stellar H$\beta$ indices observed in these galaxies
(de Jong \& Davies 1997, although their correlation 
with H$\beta$ has been challenged by 
Rampazzo et al. 1999).
Recent HI observations of low luminosity ellipticals
have detected smooth, differentially rotating 
disks of HI gas in these galaxies, sometimes extending 
right to the center 
(Oosterloo, Morganti, \& Sadler 1998) similar 
to the cold disks predicted by our models.
Evidently, star formation in large, coherent 
HI disks in low mass ellipticals 
differs from the model described here 
in which the HI gas is distributed among many isolated, 
physically distinct clouds.

Small irregular 
dusty clouds are observed in the cores of 
most ($\sim 80$ percent) bright ellipticals  
(van Dokkum \& Franx 1996). 
Centrally orbiting HII gas is also occasionally observed, 
sometimes rotating in a 
sense opposite to that of the galaxy (e.g. Bertola, 
Buson \& Zeilinger 1988).
In these cases gas has presumably been 
acquired by a merging process unrelated to cooling 
flow evolution. 
Since the dusty clouds typically 
have masses in excess of the 
critical masses for gravitational collapse $\sim 1$ 
$M_{\odot}$, luminous 
massive star formation with a normal IMF can be expected 
as well as Type II supernovae, but at very low rates. 
Normalizing to our Galaxy where there are $\sim 10^{-12}$ 
SNII yr$^{-1}$ per solar mass of molecular interstellar gas, 
the $< 10^7$ $M_{\odot}$ of cold gas in NGC 4472 
should generate $\lta 10^{-5}$ SNII yr$^{-1}$. 
Because this SNII rate is so small (and their light is masked 
by starlight in the galactic core), it is not surprising 
that SNII have escaped detection in ellipticals. 
Optical luminosity and H$\beta$ emission from normal star formation 
in dusty cores is also deeply buried in starlight.
According to the data of
Tacconi \& Young (1986), the disk of the Scd galaxy 
NGC 6946 produces $\sim 3.5 \times 10^{33}$ ergs s$^{-1}$ 
of B-band light 
per solar mass of H$_2$ gas so the $M < 10^7$ $M_{\odot}$ 
of cold gas in NGC 4472 
should generate $L_{B,d} < 3.5 \times 10^{40}$ erg s$^{-1}$. 
The dusty gas cloud in NGC 4472 has a maximum extent 
of only $r_d \sim 0.91$ arcseconds (van Dokkum \& Franx 1995) 
which is less than the ``break'' radius $r_b = 2.41''$ 
in the stellar light. 
Assuming the B-band surface brightness in NGC 4472 is 
uniform within $r_b$, the total projected luminosity produced 
by galactic stars, $L_{B,proj}(r_d) \approx 
= \pi r_d^2 \Sigma_B(r_d) \approx 4.5 \times 10^{40}$ 
erg s$^{-1}$, exceeds $L_{B,d}$.
We conclude that normal stellar formation can occur in 
accreted dusty clouds at the centers of bright ellipticals 
like NGC 4472 without easy detection.

The interesting 
possibility of observing cold gas in elliptical galaxies 
with the 6.4 keV fluorescent Fe line has been discussed by 
Churazov et al. (1998).
However, the small star-forming regions discussed here are 
completely transparent to K shell ionizing radiation at 7 keV 
(see Figure 4) so very little 6.4 keV radiation would be 
expected. 
In any case, 
galactic cooling flows of temperature $\sim 1$ keV have 
little emission above 7 keV.
Perhaps the best confirmation of our star formation model 
would be to directly observe the young, low mass stars 
and a variety of infrared techniques are available for 
this purpose 
(Kroupa \& Gilmore 1994; Joy et al. 1995; Prestwich et al. 1997).

Our conclusions can be briefly summarized as follows. 
The final stages of catastrophic cooling in galactic 
cooling flows is distributed 
among a large number of cooling sites, 
each occupying a tiny fraction of the galactic volume. 
The mass of cold gas within HII clouds 
at each site accumulates slowly, allowing 
episodic gravitational collapse of dense neutral cores. 
In the typical interstellar environment of 
giant elliptical galaxies, 
the maximum mass of gravitationally unstable regions is 
only $\lta 2$ $M_{\odot}$ (i.e. later than A5V), 
setting an absolute upper limit on the mass 
of stars that can form. 
Some optical luminosity would be produced by these 
stars even if the collapsing 
cores fragmented with a normal IMF.
Magnetic stresses and rotation 
could complicate this conclusion, but simple estimates 
indicate that magnetic fields diffuse rapidly out of  
collapsing, star-forming cores.
This method of forming low mass stars involves very little 
neutral or molecular gas and is completely consistent 
with the apparent absence of cold gas in normal 
ellipticals --  
assuming that the mass of cold gas is more accurately 
determined from radio than from X-ray observations. 
If cooling from hot interstellar gas occurs at 
large galactic radii, near $r_e$, then stars 
of larger mass $\lta 2 - 3$ $M_{\odot}$ can form. 
Such stars may help produce the H$\beta$ absorption 
features observed in elliptical galaxies 
which have been interpreted as evidence for 
a population of youthful stars of intermediate mass. 
Finally, in elliptical galaxies 
the star-forming sites are illuminated by optical 
HII line emission so the surface brightness of 
this radiation must be in accord with mass
dropout models as well as with observations of
the hot X-ray emitting interstellar gas.

\vskip.2in
We wish to thank David Boute and Neal Turner for helpful advice. 
Our work on the evolution of hot gas in elliptical galaxies 
is supported by grants NAG 5-3060 and NAG 5-8049  
from NASA and 
AST-9802994 from the National Science Foundation. 
FB is supported in part by grant MURST-Cofin 98 from the Agenzia 
Spaziale Italiana.

\clearpage

\makeatletter
\def\jnl@aj{AJ}
\ifx\revtex@jnl\jnl@aj\let\tablebreak=\nl\fi
\makeatother

\begin{deluxetable}{rrcrrr}
\footnotesize
\tablenum{1}
\tablewidth{33pc}
\tablecaption{GRAVITATIONAL STABILITY OF HI CORES\tablenotemark{a}}
\tablehead{
\colhead{$M_{HI}$} &
\colhead{$r_{HI}$} &
\colhead{$M_{HI} + M_{HII}$} &
\colhead{$r_{HII}$\tablenotemark{b}} &
\colhead{$\langle T \rangle$\tablenotemark{c}} &
\colhead{$T_m$\tablenotemark{d}} \cr
\colhead{($M_{\odot}$)} &
\colhead{(pc)} &
\colhead{($M_{\odot}$)} &
\colhead{(pc)} &
\colhead{(K)} &
\colhead{(K)} 
}
\startdata
1.0 & 0.0276 & 1.08 & 0.206 & 15.8 &  8.0  \cr
1.5 & 0.0314 & 1.58 & 0.206 & 15.4 &  9.8  \cr
2.0 & 0.0343 & 2.08 & 0.206 & 15.2 & 11.3  \cr
2.5 & 0.0368 & 2.58 & 0.206 & 15.0 & 12.7  \cr
3.0 & 0.0390 & 3.08 & 0.206 & 14.8 & 13.9  \cr
3.5 & 0.0409 & 3.58 & 0.206 & 14.6 & 15.0  \cr
4.0 & 0.0426 & 4.08 & 0.206 & 14.5 & 16.0  \cr
4.5 & 0.0442 & 4.58 & 0.206 & 14.4 & 17.0  \cr
5.0 & 0.0457 & 5.08 & 0.206 & 14.3 & 17.9  \cr
\enddata
\tablenotetext{a}{External pressure $P = 2 \times 10^{-10}$ dynes, 
${\cal J}_{uv} = 1 \times 10^8$ cm$^{-2}$ s$^{-1}$ 
ster$^{-1}$, and $T_{HII} = 10^4$ K for
all models.}
\tablenotetext{b}{$r_{HII}$ is the outer radius of the entire cloud,
most of which is ionized.}
\tablenotetext{c}{Mass-weighted average temperature in HI core.}
\tablenotetext{d}{Minimum temperature for pressure-supported 
HI cloud cores.}
\end{deluxetable}

\clearpage

\vskip.1in
\figcaption[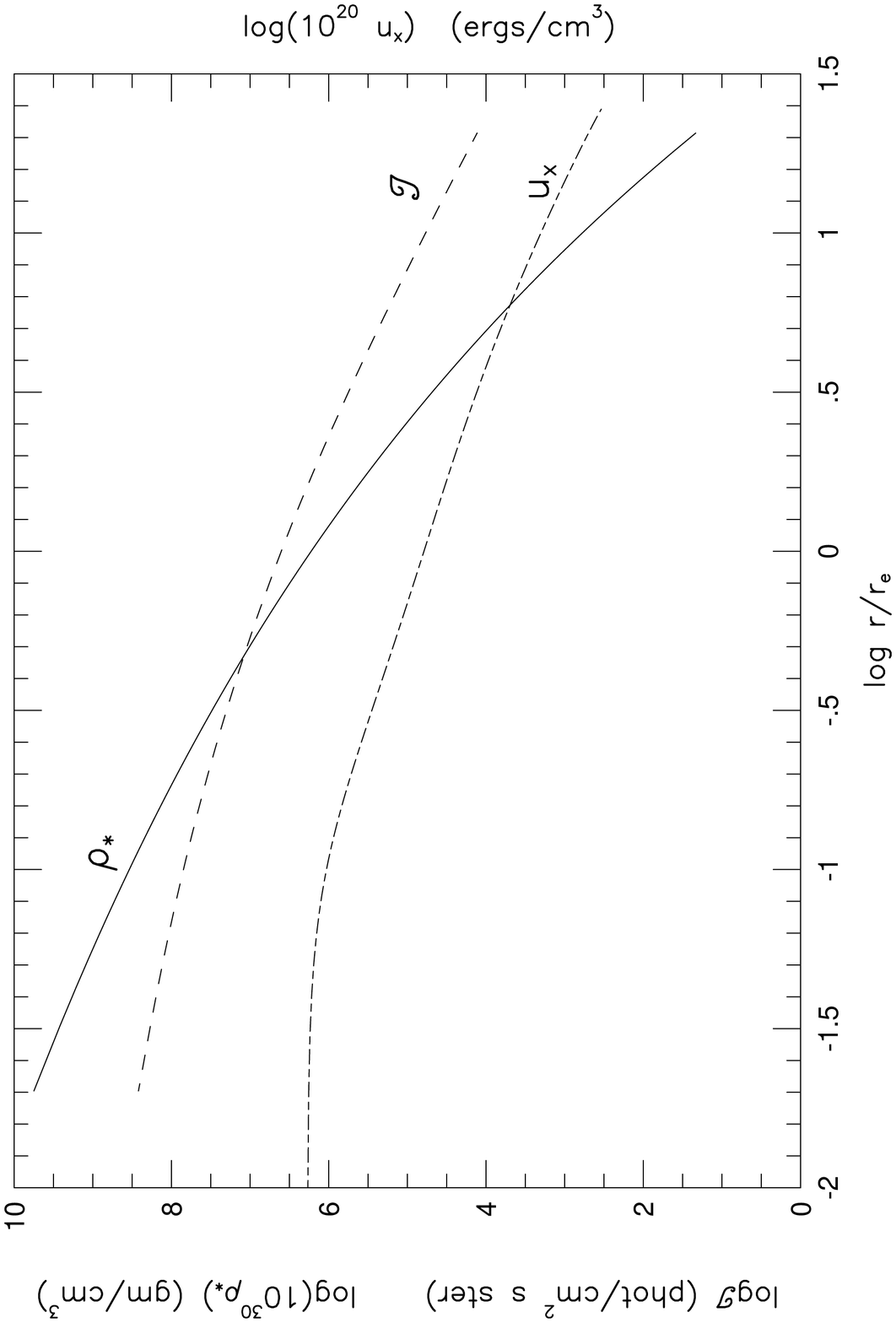]{
Radial variation of stellar density $\rho_*(r/r_e)$ in NGC 4472 
({\it solid line}) and corresponding mean photon intensity 
in the H-ionizing continuum, 
${\cal J}(r/r_e)$ photons cm$^{-2}$ s$^{-1}$ ster$^{-1}$ 
({\it dashed line}).
The bolometric energy density of thermal X-radiation from 
the hot interstellar gas $u_x(r/r_e)$ is shown for 
comparison ({\it long-short dashed line}).
The radius is normalized to the effective radius 
$r_e = 1.73$ arc minutes or $8.57$ kpc at $D = 17$ Mpc.
\label{fig1}
}

\vskip.1in
\figcaption[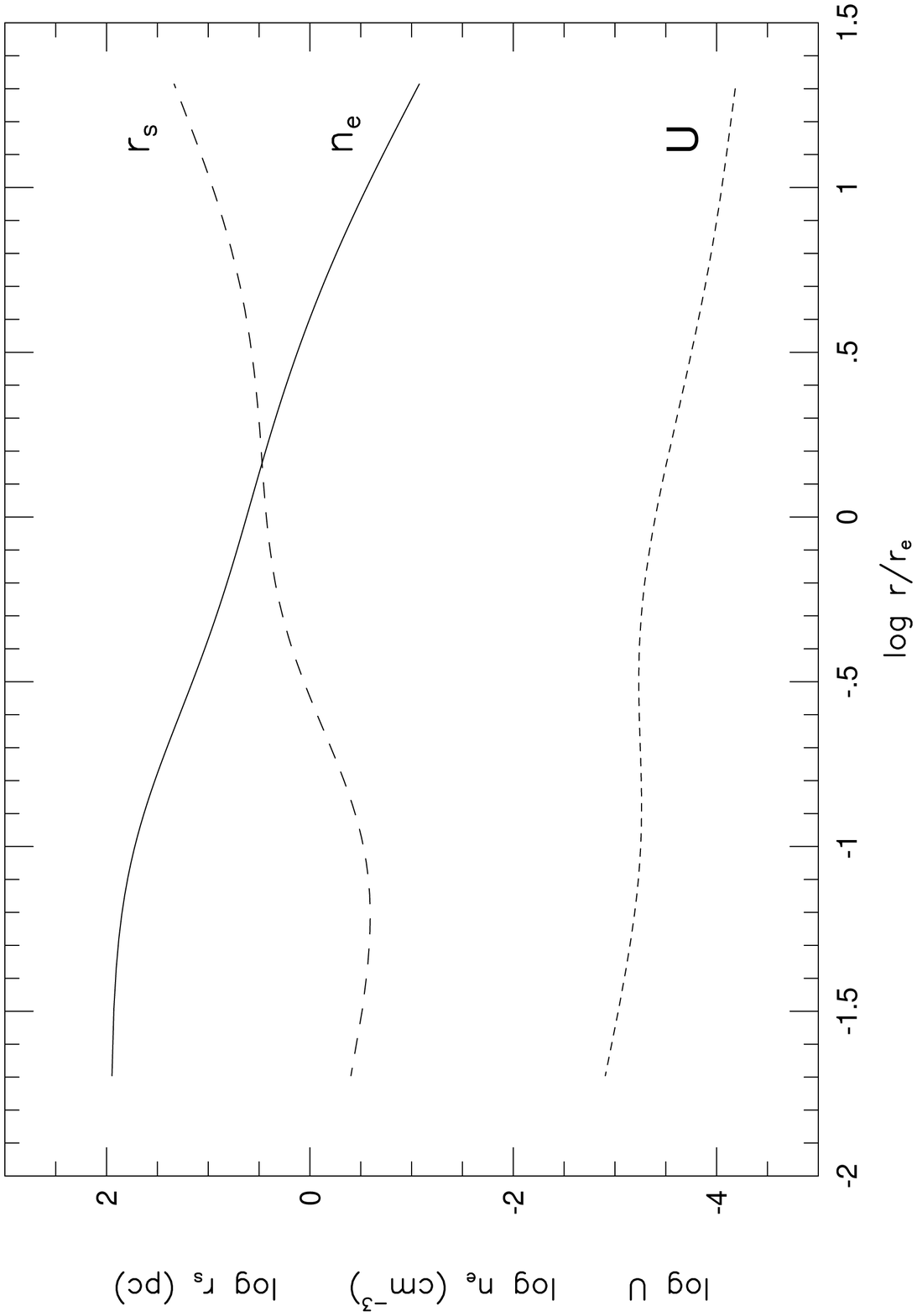]{
Radial variation of HII gas density 
$n_e$ ({\it solid line}), HII Stromgren radius 
$r_s$ ({\it long dashed line}), 
and ionization parameter $U = n_{iph}/n_e$ 
({\it short dashed line}) as functions of galactic 
radius in NGC 4472. 
The radius is normalized to the effective radius 
$r_e = 1.73$ arc minutes or $8.57$ kpc at $d = 17$ Mpc.
\label{fig2}
}

\vskip.1in
\figcaption[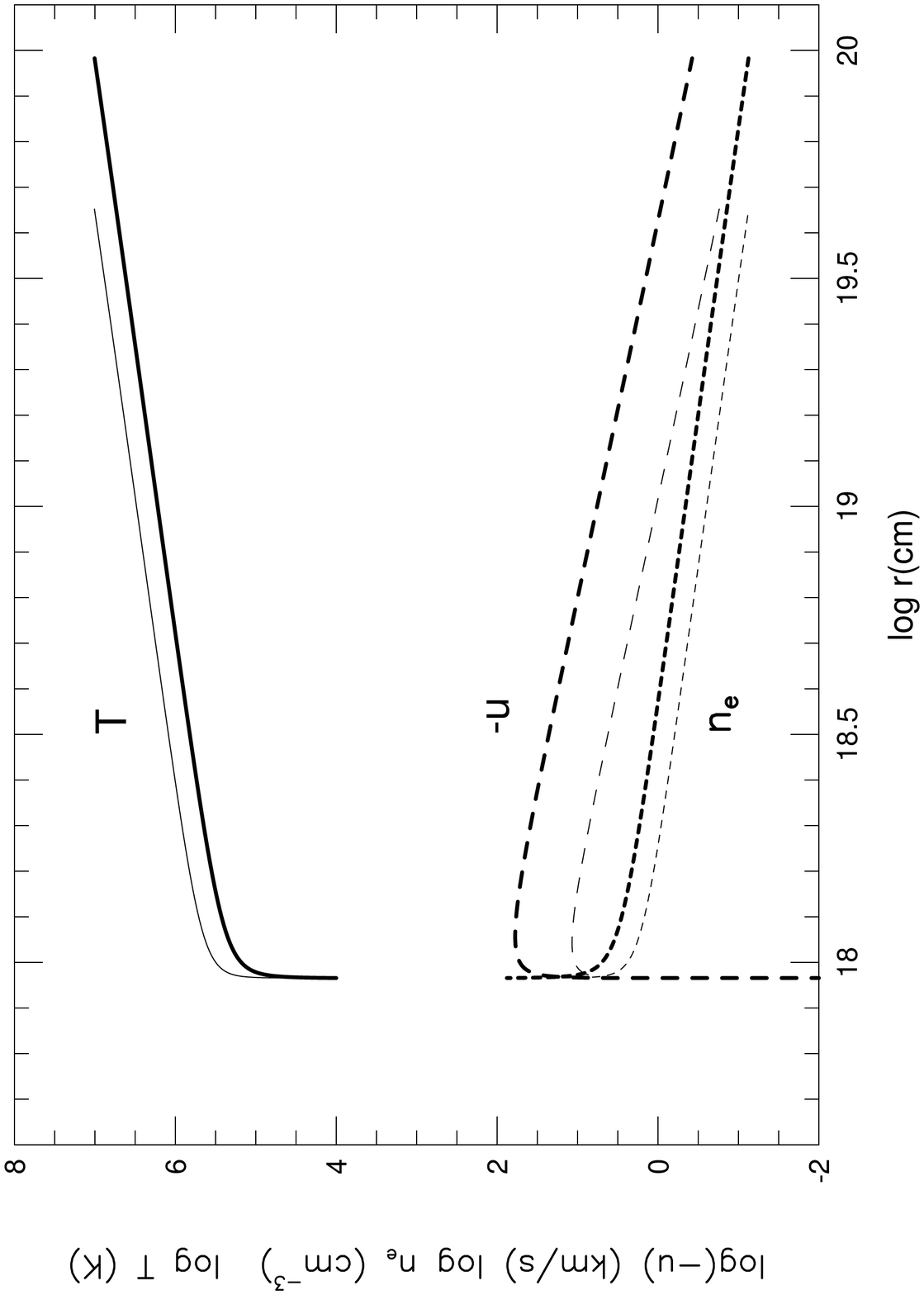]{
Steady state spherical gas flow surrounding 
HII gas of radius $0.3$ pc at a 
typical galactic cooling site.
Gas temperature, velocity and density are shown 
with solid, long-dashed and short-dashed lines 
respectively.
Heavy lines correspond to 
${\dot m} = 10^{-5}$ $M_{\odot}$ yr$^{-1}$
and light lines to 
$10^{-6}$ $M_{\odot}$ yr$^{-1}$.
\label{fig3}
}

\vskip.1in
\figcaption[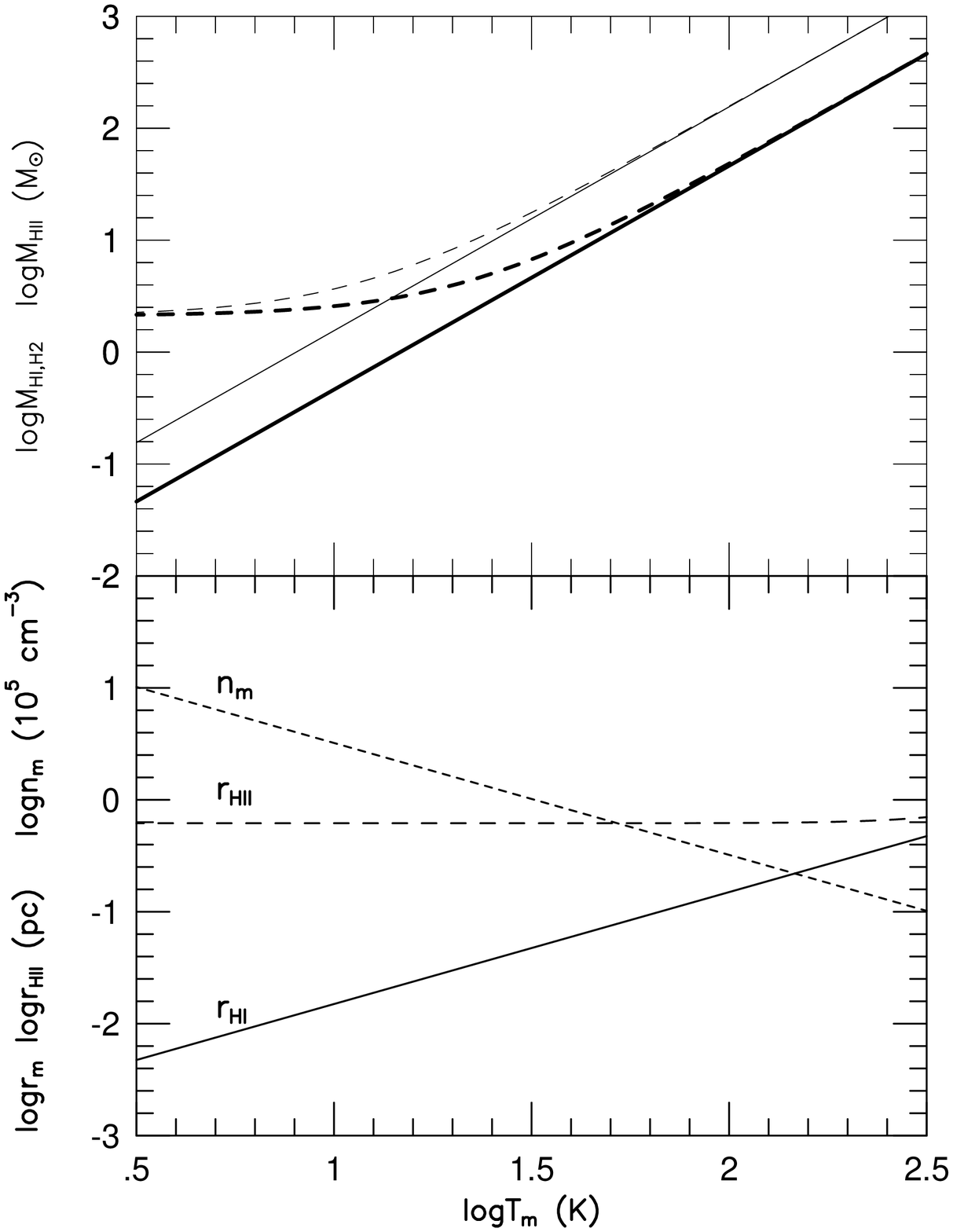]{
Properties of uniform 
clouds that are marginally stable to 
gravitational collapse in an environment with 
external pressure $P = 2 \times 10^{-10}$ dynes. 
{\it Upper panel:} Masses of neutral 
({\it light solid line}) and molecular
({\it heavy solid line}) cores and including their 
corresponding HII envelopes ({\it dashed lines})
as functions of the critical core temperature 
$T_m$.
{\it Lower panel:} Density in core HI gas 
$n$ ({\it short dashed line}), radii of 
HI core ({\it solid line}) and HII envelope 
({\it long dashed line}) as functions of 
the critical core temperature $T_m$.
\label{fig4}
}

\vskip.1in
\figcaption[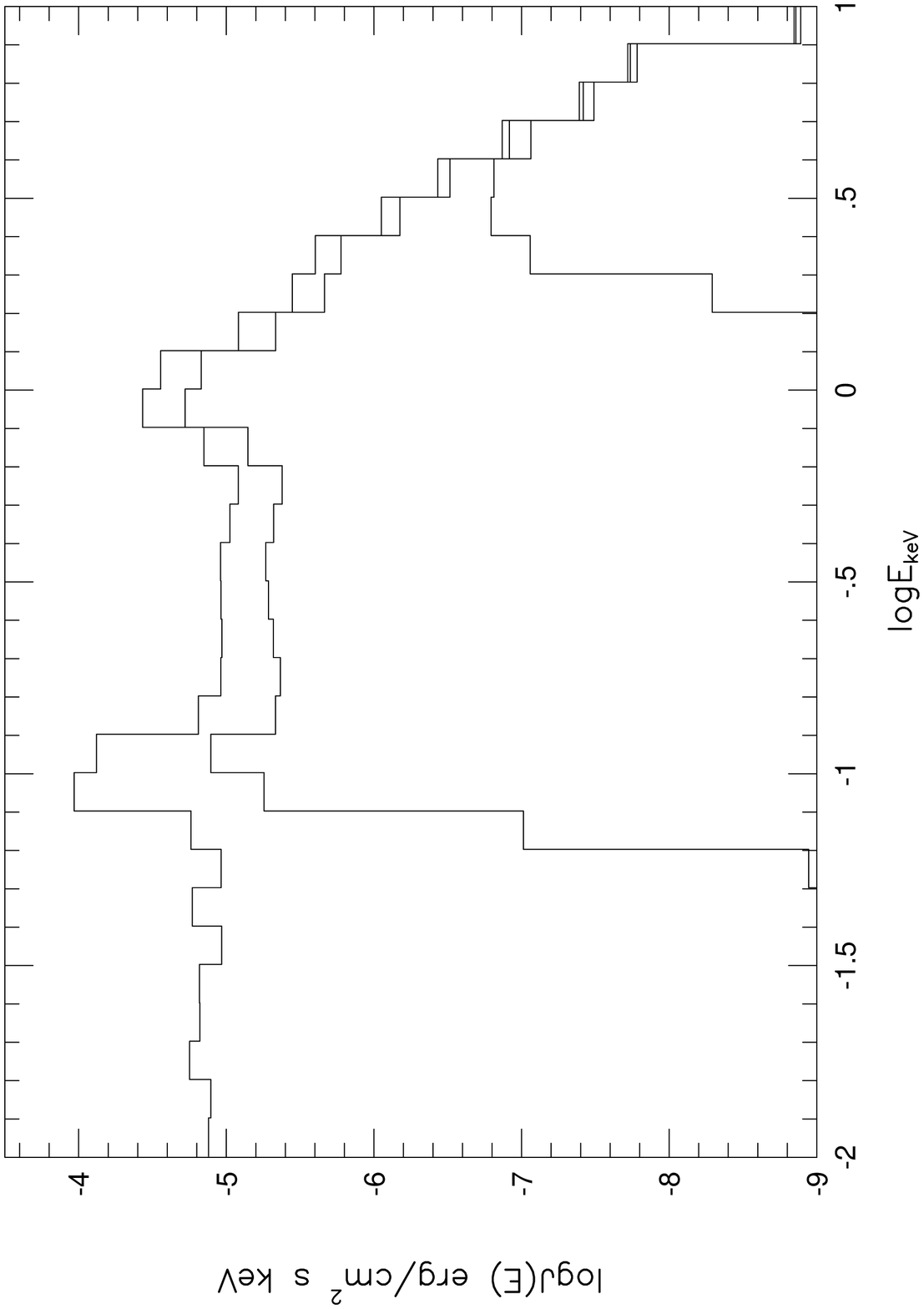]{
X-ray spectrum at three radii within an 
HII-HI cloud having core mass 
$M_{HI} = 3.5$ $M_{\odot}$. 
The spectrum is shown at three radii:
$r_{HII} = 0.206$ pc ({\it upper}),
$r_{HI} = 0.0409$ pc ({\it middle}), and 
$r = 0$ pc ({\it lower}).
\label{fig5}
}

\vskip.1in
\figcaption[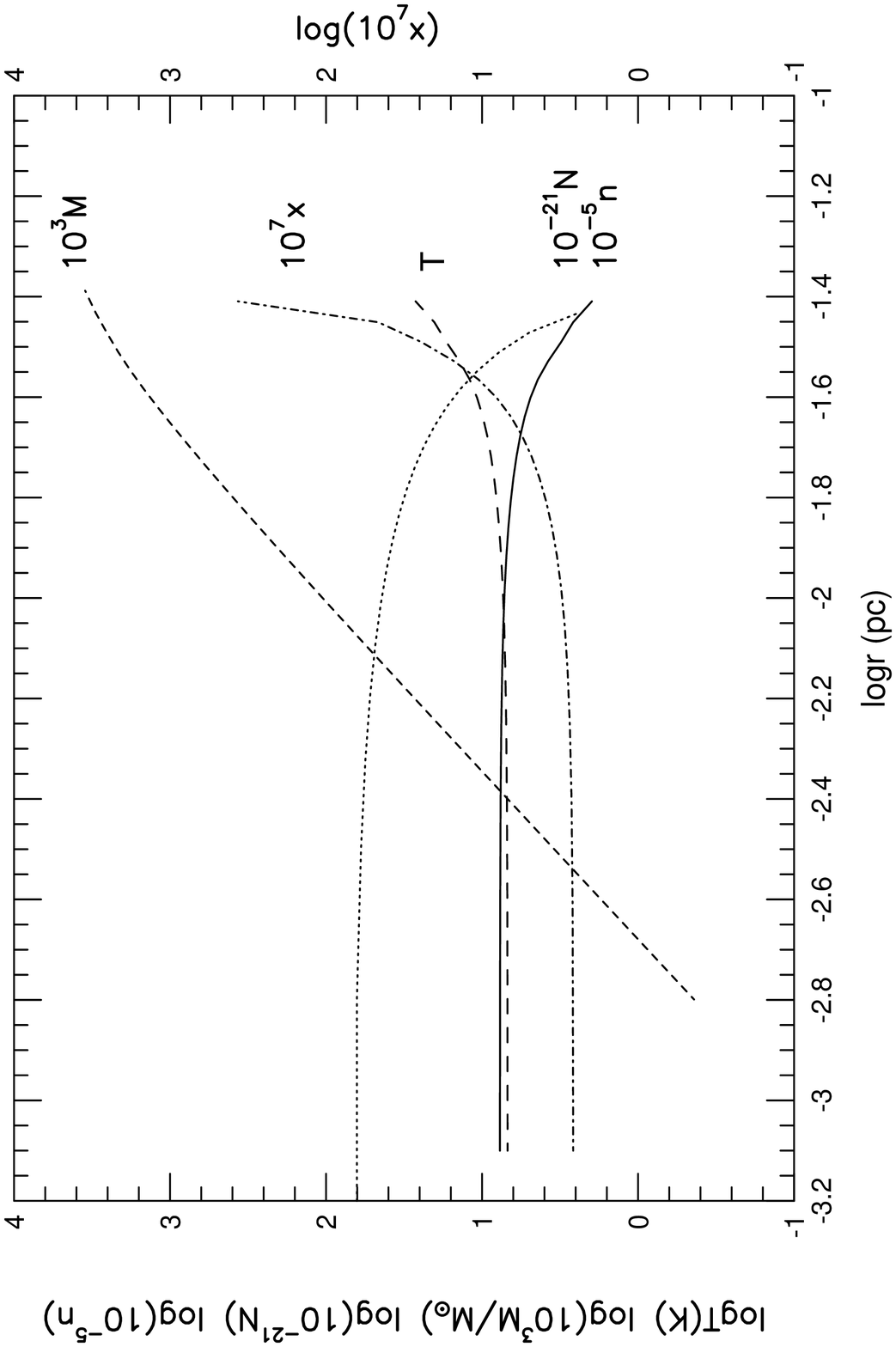]{
Equilibrium structure of an HI core of mass 
$M_{HI} = 3.5$ $M_{\odot}$.
Shown as functions of radius 
are the temperature $T$ 
({\it long dashed line}),
HI density $n$
({\it solid line}),
column density $N$
({\it dotted line}),
degree of ionization $x$
({\it dash-dot line}),
and 
the mass $M$
({\it short dashed line}).
The pressure $P = 8 \times 10^{-10}$ dynes 
is uniform.
\label{fig6}
}

\vskip.1in
\figcaption[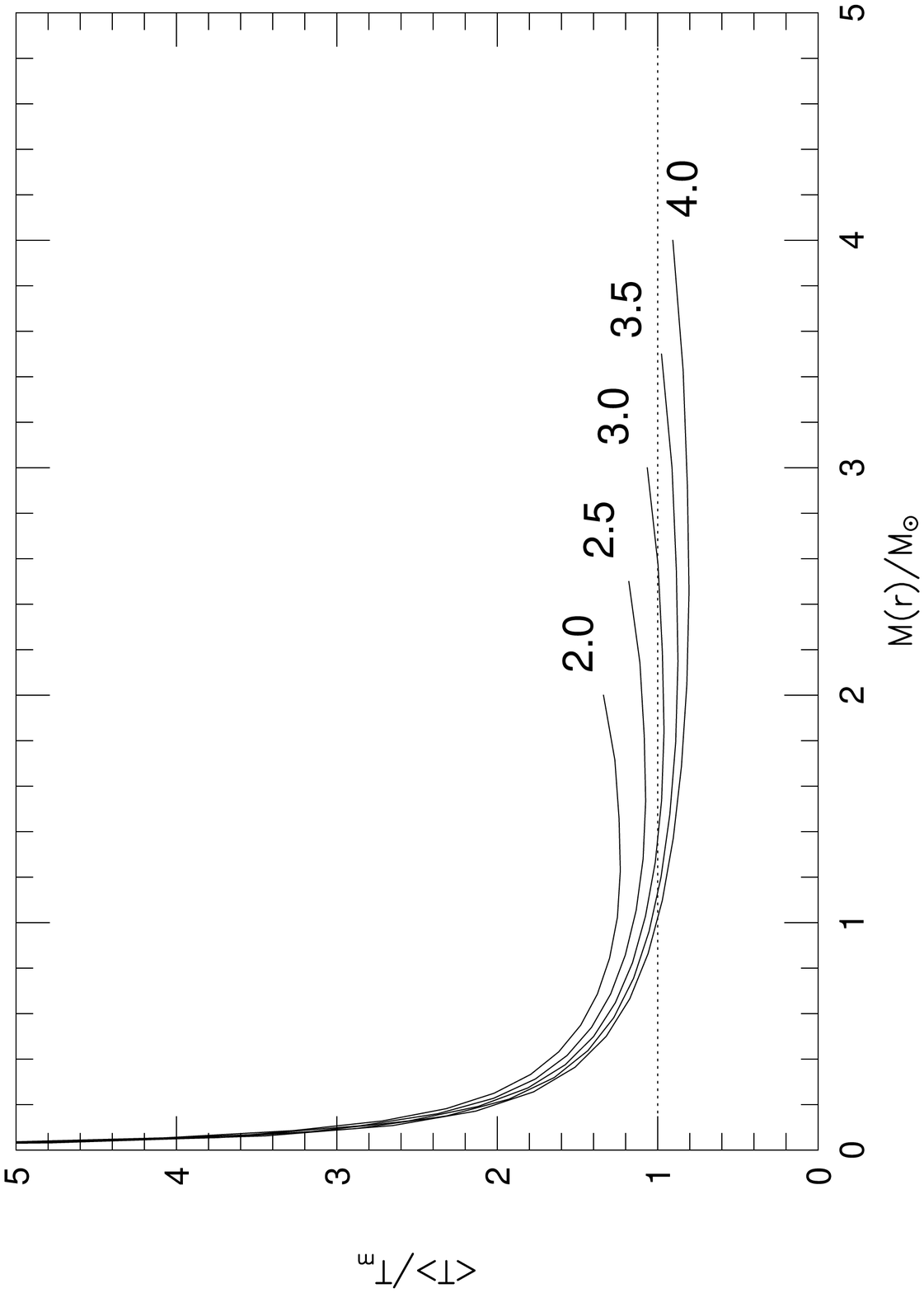]{
Variation of the ratio of true mean temperatures 
$\langle T \rangle$ within mass $M(r)$ to the 
minimum temperature for gravitational support $T_m$ 
for five HI cores of different total mass. 
Regions for which $\langle T \rangle/T_m < 1$ cannot  
support against gravity and collapse ensues.
Each curve is labeled with the total HI core mass 
in solar units.
\label{fig7}
}

\vskip.1in
\figcaption[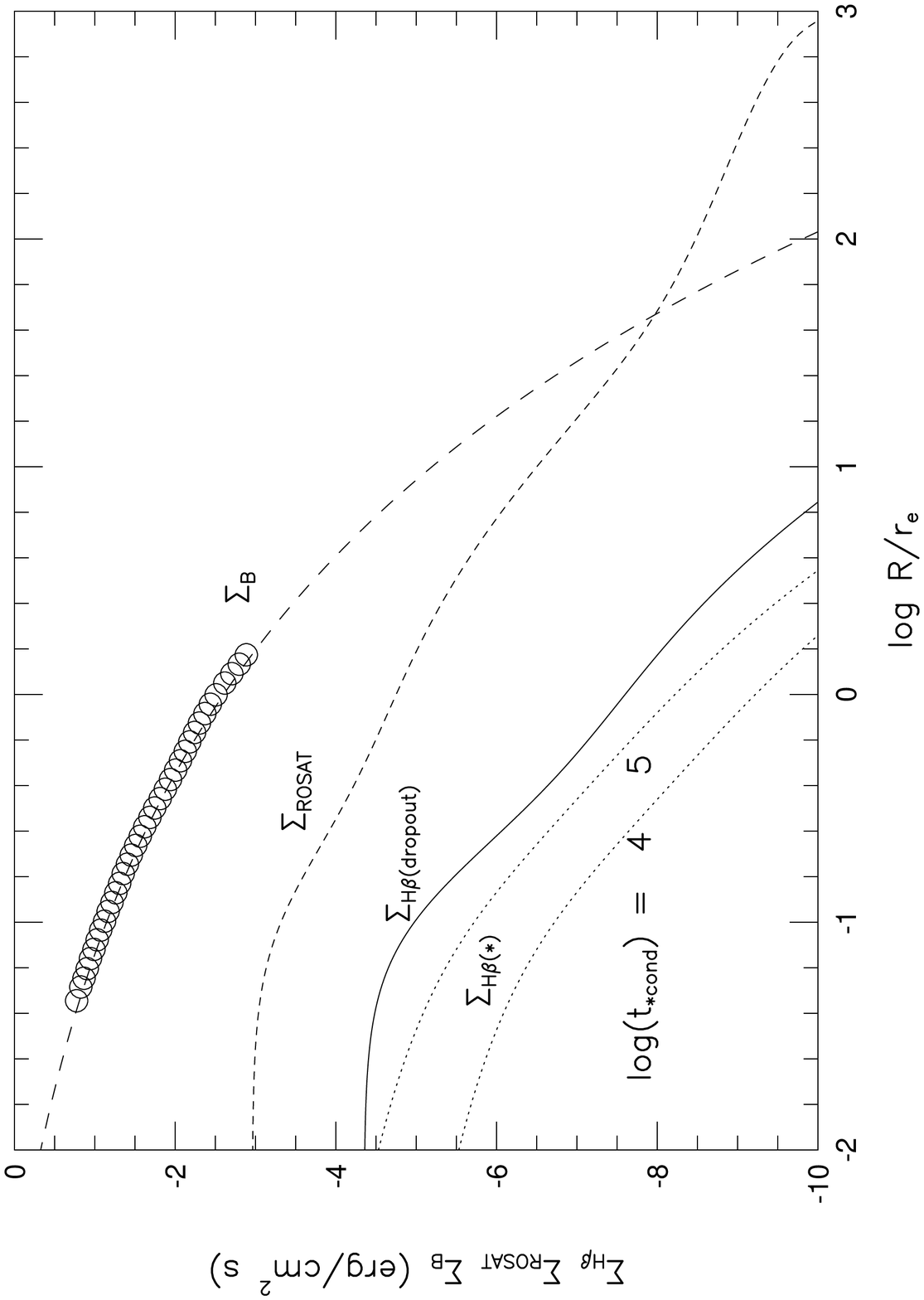]{
Optical and X-ray surface brightness distributions in 
NGC 4472:
{\it long dashed line:} B-band 
surface brightness $\Sigma_B$ with photometric observations of  
Peletier et al. (1990) shown with open circles;
{\it short dashed line:} 
ROSAT band surface brightness $\Sigma_{ROSAT}$
calculated from adopted 
density and temperature distributions; 
{\it solid line:} H$\beta$ surface brightness 
$\Sigma_{H\beta(dropout)}$ with $q = 1$ dropout model;
{\it dotted lines:} H$\beta$ surface brightness
$\Sigma_{H\beta(*)}$ of gas ejected from stars 
for two assumed HII lifetimes, $t_{*,cond} = 10^4$ and $10^5$
years.
\label{fig8}
}

\end{document}